\documentclass[aps,twocolumn,showpacs,preprintnumbers,nofootinbib,prd,superscriptaddress,groupedaddress,10pt]{revtex4-1}

\makeatletter
\def\l@subsubsection#1#2{}
\def\l@subsubsubsection#1#2{}
\makeatother

\usepackage{graphicx,amssymb,amsmath,amsthm,amsfonts,epsfig,epsf,fixmath}
\usepackage[usenames]{color}
\usepackage{epstopdf}

\usepackage{aas_macros}
\usepackage{bm}
\usepackage{dcolumn}
\usepackage{latexsym}
\usepackage{rotating}
\usepackage{longtable}

\usepackage{enumerate}
\usepackage{tensor,multirow}
\usepackage{url}
\usepackage[linktocpage]{hyperref}

\newcommand{\PS}{{\sc ParSpec }}

\begin{document}
\title{
Parametrized ringdown spin expansion coefficients:\\a data-analysis 
framework for black-hole spectroscopy with multiple events}
\author{
Andrea Maselli$^1$,
Paolo Pani$^{1}$,
Leonardo Gualtieri$^1$,
Emanuele Berti$^2$}

\affiliation{$^{1}$ Dipartimento di Fisica, ``Sapienza'' Universit\`a di Roma, Piazzale 
Aldo Moro 5, 00185, Roma, Italy}


\affiliation{$^{2}$ Department of Physics and Astronomy, John Hopkins University, Baltimore, MD 21218 USA}

\begin{abstract} 
Black-hole spectroscopy is arguably the most promising tool to test 
gravity in extreme regimes and to probe the ultimate nature of black 
holes with unparalleled precision. These tests are currently limited by 
the lack of a ringdown parametrization that is both robust and accurate.
We develop an observable-based parametrization of the ringdown of 
spinning black holes beyond general relativity, which we dub \PS 
(Parametrized Ringdown Spin Expansion Coefficients).  This approach 
is perturbative in the spin, but it can be made arbitrarily precise (at least 
in principle) through a high-order expansion. It requires ${\cal O}(10)$ 
ringdown detections, which should be routinely available with the planned 
space mission LISA and with third-generation ground-based detectors.
We provide a preliminary analysis of the projected bounds on parametrized 
ringdown parameters with LISA and with the Einstein Telescope, and discuss 
extensions of our model that can be straightforwardly included in the future.
\end{abstract}

\maketitle

\section{Introduction}
Atomic spectroscopy revolutionized the quantum description of atomic interactions and paved the wave for quantum 
electrodynamics, through the precise measurements of the energy levels of the hydrogen atom~\cite{LambShift}.
Black-hole~(BH)
spectroscopy~\cite{1980ApJ...239..292D,Dreyer:2003bv,Berti:2005ys}
--~i.e., the measurement of the quasinormal modes~(QNMs) of a
BH~\cite{Vishveshwara:1970cc,Chandra,Kokkotas:1999bd,Ferrari:2007dd,Berti:2009kk,Konoplya:2011qq}
through gravitational wave~(GW) ringdown observations~-- may play a
similar major role to probe the gravitational interaction and
fundamental physics in extreme
conditions~\cite{Berti:2015itd,Barack:2018yly,Berti:2018vdi,Berti:2019xgr,Sathyaprakash:2019yqt}.

The post-merger ringdown signal from a remnant BH can be modeled as a
superposition of damped
sinusoids~\cite{Kokkotas:1999bd,Ferrari:2007dd,Berti:2009kk}, each
defined by an oscillation frequency $\omega$ and a damping time
$\tau$. Owing to the BH uniqueness and no-hair
theorems~\cite{Carter71,Hawking:1973uf,Robinson,Cardoso:2016ryw}, the
entire QNM spectrum of a spinning~(Kerr) BH in general relativity~(GR)
is completely determined by the mass $M$ and spin $J=\chi M^2$ of the
BH. (We use $G=c=1$ throughout.)
Thus, measuring one frequency and damping time allows us to infer the
mass and spin of a merger remnant from the ringdown signal only,
whereas measuring more than two quantities (i.e., also subdominant
modes) provides multiple independent null-hypothesis tests of
GR~\cite{Dreyer:2003bv,Berti:2005ys,Berti:2007zu,Gossan:2011ha,Meidam:2014jpa,Bhagwat:2017tkm,Baibhav:2017jhs,
Baibhav:2018rfk,Brito:2018rfr,Carullo:2018sfu}.
In addition, measuring also the amplitudes of multiple modes provides information about the intrinsic parameters of the 
progenitor binary~\cite{Hughes:2019zmt,Apte:2019txp,Lim:2019xrb}.
These tests require high signal-to-noise ratio (SNR) in the
ringdown~\cite{Berti:2005ys,Berti:2007zu} and will become routinely
available with the space mission LISA~\cite{Audley:2017drz} and with
third-generation~(3G) ground-based GW detectors (such as the proposed
Einstein Telescope~(ET)~\cite{Punturo:2010zz} and Cosmic
Explorer~\cite{Evans:2016mbw}), which are expected to detect several
ringdown events per year with SNR in the hundreds to thousands, even
from sources at cosmological distance~\cite{Berti:2016lat}.

The LIGO/Virgo Collaboration checked that the full
inspiral-merger-ringdown waveform is consistent with GR by analyzing
separately the lower-frequency signal emitted during the inspiral
phase and the higher-frequency signal emitted during the late
inspiral, merger and ringdown of the first event,
GW150914~\cite{TheLIGOScientific:2016src}. Separately fitting each of
these signals to GR-based templates leads to two independent estimates
of the mass and dimensionless spin of the remnant BH. An extension of
this analysis to seven selected binary BH events found that the two
estimates are compatible with each other within statistical errors of
order $30\%$ (see Fig. 2 in~\cite{LIGOScientific:2019fpa}). Recent
work tried to better quantify the contribution of additional overtones
to the high-frequency signal for GW150914. Adding at least one
overtone is necessary to obtain ringdown estimates of the mass and
spin of the remnant which are in agreement (at the $\sim 20\%$ level)
with the values inferred by fitting the entire signal within
GR~\cite{Giesler:2019uxc,Isi:2019aib} (see
also~\cite{Berti:2007zu,Baibhav:2017jhs,Bhagwat:2019dtm}).

Going beyond these consistency tests requires modeling the BH ringdown
beyond GR, for instance to perform a Bayesian model selection between
GR and any proposed extension of the theory.  This is a challenging
task and, despite recent
progress~\cite{Barausse:2014tra,Glampedakis:2017dvb,Glampedakis:2017cgd,Tattersall:2017erk,Franciolini:2018uyq,Cardoso:2019mqo,McManus:2019ulj,Glampedakis:2019dqh},
all current attempts have significant limitations: they are based on
particular classes of theories, they use geometric-optics
approximations for the QNMs, or they neglect the spin of the remnant.

Working in the nonrotating limit is a major limitation, since the
final spin of the merger remnant is typically
high~\cite{Buonanno:2006ui,Berti:2007fi,Berti:2008af,Hofmann:2016yih}. Including
spin in current approaches is challenging, especially because the
geometry of spinning BHs beyond GR is known only perturbatively or
numerically (see
e.g.~\cite{Pani:2011gy,Kleihaus:2011tg,Ayzenberg:2014aka,Maselli:2015tta,Barausse:2015frm,Herdeiro:2016tmi,Cunha:2019dwb}
for specific examples
and~\cite{Berti:2015itd,Herdeiro:2015waa,Yagi:2016jml} for reviews),
which makes it very hard to compute the QNMs. In addition, there is in
general no analog of the Teukolsky
equation~\cite{Teukolsky:1972my,Teukolsky:1973ha,Chandra} beyond
GR. In general the perturbation equations are not
separable~\cite{Pani:2013pma}, and this requires the solution of an
elliptic system of partial differential equations~\cite{Dias:2015wqa}
or the extraction of QNM frequencies from numerical-relativity
simulations of BH mergers~\cite{Okounkova:2017yby,Witek:2018dmd,Okounkova:2019dfo,Okounkova:2019zep}.

In contrast to these major technical limitations, modeling the BH QNMs beyond GR is remarkably straightforward.
In any extension of GR, the QNMs of a BH can be parametrized as~\cite{Gossan:2011ha,Meidam:2014jpa,Carullo:2018sfu}
\begin{eqnarray}
  \omega &=&\omega^{\rm Kerr}+\delta \omega \,, \label{wR0}\\
  \tau   &=&\tau^{\rm Kerr}  +\delta \tau   \,, \label{tau0}
\end{eqnarray}
where the frequency $\omega^{\rm Kerr}$ and damping time $\tau^{\rm Kerr}$ depend only on $M$ and $\chi$, whereas $\delta\omega$ and $\delta\tau$ 
are generic deviations.
We consider a modified ringdown which deviates \emph{perturbatively} from the Kerr case in GR, i.e. $\delta\omega\ll 
\omega^{\rm Kerr}$ and $\delta\tau\ll 
\tau^{\rm Kerr}$. These departures can be 
due to extra charges, a modified theory of gravity, environmental effects, etcetera, and we wish to develop a 
generic framework that can accommodate various special cases.
GR corrections might affect the ringdown in two ways: by predicting a
spinning BH other than
Kerr~\cite{Kleihaus:2011tg,Ayzenberg:2014aka,Barausse:2015frm,Cardoso:2018ptl,Yagi:2016jml,Berti:2015itd,Barack:2018yly,Sathyaprakash:2019yqt},
or (even if GR BHs are still solutions of the theory) by affecting the
dynamics of the
perturbations~\cite{Barausse:2008xv,Molina:2010fb,Tattersall:2017erk,Tattersall:2018map,Tattersall:2018nve,Tattersall:2019pvx}. In
both cases, the ringdown modes will acquire corrections proportional
to the fundamental coupling constant(s) of the theory.  There may be
new classes of modes associated to extra polarizations, but they are
unlikely to be significantly excited for GR deviations small enough to
be compatible with existing
observations~\cite{Molina:2010fb,Barausse:2014tra,Blazquez-Salcedo:2016enn}.
For this reason they will not be considered in our analysis.

The above discussion suggests that a case-by-case analysis is impractical, and that parametrizing directly the observables (i.e.,
frequencies and damping times) is the most efficient way to
perform ringdown tests (see e.g.~\cite{Meidam:2014jpa,Carullo:2018sfu,Cardoso:2019mqo,McManus:2019ulj} for work
in this direction).  Similar observable-based approaches have been
very successful to model weak-field effects~\cite{Will:2014kxa} and
the inspiral~\cite{Yunes:2009ke,Agathos:2013upa}.

In this paper we develop a scheme based on ``Parametrized Ringdown
Spin Expansion Coefficients''~({\sc ParSpec}) which differs from
related hierarchical approaches~\cite{Isi:2019asy} and mode-stacking
proposals~\cite{Yang:2017zxs}. Its salients features are as follows:

\begin{itemize}
\item[1)] We expand the spectrum in a bivariate series in terms of
  the fundamental parameters (mass and spin) characterizing BH
  dynamics in GR.
\item[2)] The expansion parameters take into account the fact that
  modifications of GR are suppressed by a (possibly dimensionful)
  coupling constant.
\item[3)] Bayesian inference allows us to identify the most easily
  measurable expansion coefficients. By combining several
  observations, we can in principle map the deviation parameters to
  specific modified theories of gravity for which QNM spectra may be
  available. Since the third LIGO/Virgo observing run has been
  detecting BH mergers on a weekly basis, this approach holds the
  promise of allowing us to constrain several parameters (or identify
  deviations) as soon as the typical SNR of the observations becomes
  large enough.
\end{itemize}
  
The plan of the paper is as follows. In Section~\ref{sec:framework} we
describe our parametrized framework. In Section~\ref{sec:stat} we
illustrate the potential of the method by performing a statistical
analysis on a representative catalog of merger events with LISA and 3G
Earth-based detectors. In Sections~\ref{sec:future} and
~\ref{sec:conclusions} we compare our framework with previous work and
discuss directions for future research.

\section{PARSPEC framework}\label{sec:framework}

Let us assume $i=1,\,\dots,\,N$ independent ringdown detections, for 
which $q$ QNMs are measured.
In general $q$ depends on the source $i$, but for simplicity we shall
consider a subset of all $N_T$ merger events for which the {\em same}
number $q$ of QNMs passes a certain SNR threshold.
Therefore $N$ is (in general) smaller than $N_T$, but this is not a major 
limitation, given the high event rates expected for future detectors.

In terms of a standard spheroidal-harmonics decomposition~\cite{Berti:2009kk}, and depending on the intrinsic parameters
of the progenitor binary (mass ratio and spins), typically the most excited QNMs\,\footnote{The QNMs are identified
  by three integer numbers: the angular momentum number $l$, the azimuthal number $m\in[-l,l]$, and the overtone number,
  which we set to zero in this paper, i.e. we only consider fundamental modes.  For ease of notation we leave these
  indices implicit, i.e. $\omega^{(J)}\equiv \omega^{(0lm)}$, where $J$ is an index that labels the mode.}
are the \emph{fundamental} modes with $l=m=2$, $l=m=3$, and $l=2$, $m=1$.
For simplicity we will assume the subdominant mode to be $l=m=3$ for
all $N_T$ sources; this assumption will be justified below.
For a given $(l,m)$, the overtones are in general relevant for
parameter
estimation~\cite{Leaver:1986gd,Berti:2006wq,London:2014cma,Baibhav:2017jhs,Giesler:2019uxc,Isi:2019aib}. However,
the frequencies of different overtones are very similar and hard to
resolve~\cite{Berti:2005ys,Bhagwat:2019dtm}, and therefore it is hard to use
them for direct BH spectroscopy. For this reason, in this paper we will not
consider overtones.

\begin{table*}[t]
 \begin{tabular}{ccccccc}
  \hline
\multicolumn{1}{c}{}  & \multicolumn{2}{c}{$l=2, m=2$}& \multicolumn{2}{c}{$l=3, m=3$}& \multicolumn{2}{c}{$l=2, m=1$}
  \\
  \hline
  $n$	& $w^{(n)}$	& $t^{(n)}$  	 & $w^{(n)}$	& $t^{(n)}$ 	& $w^{(n)}$	& $t^{(n)}$     \\
  \hline
  $0$	& $0.3737$	& $11.2407$	 & $0.5994$	& $10.7871$	& $0.3737$	& $11.2407$	\\
  $1$	& $0.1258$	& $0.2522$	 & $0.2021$	& $0.2276$	& $0.0629$	& $0.1261$	\\
  $2$	& $0.0717$	& $0.6649$	 & $0.1072$	& $0.8238$	& $0.0449$	& $0.7710$	\\
  $3$	& $0.0480$	& $0.5866$	 & $0.0689$	& $0.7353$	& $0.0218$	& $0.3821$	\\
  $4$	& $0.0350$	& $0.5797$	 & $0.0491$	& $0.0685$	& $0.0163$	& $0.5565$	\\
  \hline 
 \end{tabular}
 \caption{Coefficients of the spin expansion for the QNMs $l=m=2$, $l=m=3$ and $l=2$, $m=1$ of a Kerr BH in GR, obtained from the numerical data in \cite{ringdown}.}\label{tab:Kerr}
\end{table*}

Rather than considering the corrections in Eqs.~\eqref{wR0} and
\eqref{tau0} as independent parameters, it is sensible and convenient
to reduce the dimensionality of the parameter space by performing a
spin expansion. To this aim, we parametrize each mode of the $i$-th source as
\begin{eqnarray}
 \omega_i^{(J)} &=&\frac{1}{M_i} \sum_{n=0}^D  \chi^n_i w^{(n)}_J\left(1+\gamma_i\delta 
w^{(n)}_J\right) \,,\label{model1}\\
 \tau_i^{(J)} &=& M_i \sum_{n=0}^D \chi^n_i t^{(n)}_J\left(1+\gamma_i\delta t^{(n)}_J\right) \,,\label{model2}
\end{eqnarray}
where $J=1,2,...,q$ labels the mode;
$M_i$ and $\chi_i\ll1$ are the detector-frame mass
and spin of the $i$-th source, \emph{both measured assuming GR} (see below); $D$ is the order of the spin expansion; 
$w^{(n)}_J$ and $t^{(n)}_J$ are the dimensionless coefficients of the spin expansion for a Kerr BH
in GR (provided in Table~\ref{tab:Kerr} for a few representative
modes); $\gamma_i$ are dimensionless coupling constants, which can depend on
the source $i$ -- see Eq.~\eqref{gammai} below -- but do not depend on the
specific QNM; and $\delta w^{(n)}_J$ and $\delta t^{(n)}_J$ are
``beyond-Kerr'' corrections to the QNM frequencies.  Crucially, the
latter are universal dimensionless numbers that do not depend on the
source. Any possible source dependence is parametrized through
$\gamma_i$, as discussed below.

As customary in parametrized approaches, we focus on perturbative
corrections by assuming $\gamma_i \delta w^{(n)}\ll1$,
$\gamma_i \delta t^{(n)}\ll1$, and GR is recovered in the limit
$\gamma_i\to0$.

We remark that in the parametrization~\eqref{model1}, \eqref{model2}, $M_i$ and $\chi_i$ are the BH masses (in the
detector frame) and spins extracted assuming GR. In a non-GR theory, these are generally different from the actual BH
masses and spins, $\bar M_i$, $\bar\chi_i$ (see Appendix~\ref{app:framework}). Therefore, the coefficients $\gamma_i$,
$\delta w^{(n)}_J$ and $\delta t^{(n)}_J$ also include the shift between $M_i$, $\chi_i$ and the physical masses and
spins. Since $M_i$ and $\chi_i$ refer to the GR values of the detector-frame mass and spin of the $i$-th source, they
can be computed either from the full inspiral-merger-ringdown waveform within GR or from a measurement of the $l=m=2$ 
mode with a standard GR ringdown template (without any assumption on the
luminosity distance of the source). The remaining parameters in Eqs.~\eqref{model1} and \eqref{model2} are discussed
below for various special cases.

\subsection{Special cases}

\noindent {\bf Case~I: scale-free corrections.}
The simplest parametrized beyond-Kerr correction corresponds to having
$\gamma_i=\alpha$ for all sources, where $\alpha$ is a {\em
  dimensionless} coupling constant. Then $\alpha$ can be re-absorbed
within $\delta w^{(n)}$ and $\delta t^{(n)}$, and (assuming that $M_i$
and $\chi_i$ are known within some parameter estimation uncertainty)
we can parametrize the QNM spectrum beyond GR with
\begin{equation}
 {\cal P}=2(D+1) q \label{param}
\end{equation}
parameters, where $2(D+1) q$ is the total number of $\delta w^{(n)}$
and $\delta t^{(n)}$ parameters required if we consider $q$ modes up
to order $D$ in the spin expansion.

\noindent {\bf Case~II: single dimensionful coupling.}
A more general model consists of a single fundamental, {\em dimensionful} coupling constant $\alpha$ (the extension to
multiple coupling constants is straightforward). Without loss of generality, we assume that $\alpha$ has mass dimensions
$[\alpha]={\hat M}^p$,
where $p$ is fixed by the theory (for $p=0$ we recover Case~I above). Here ${\hat M}$ is the typical
mass/length scale in the problem, which for a BH coincides with its mass {\em in the source frame} $M^{\rm s}$,
as measured within GR (see Appendix~\ref{app:framework} for a discussion).
In this case, since the coefficients $\gamma_i$ are linear in the
coupling, to leading order in our perturbative scheme
\begin{equation}
 \gamma_i =\frac{\alpha}{(M_i^{\rm s})^p}=\frac{\alpha(1+z_i)^p}{M^p_i}\,, \label{gammai}
\end{equation}
are small dimensionless couplings that depend on the theory, on the
source mass in the detector frame $M_i$, and on the source redshift
$z_i$. The redshift can be estimated from the luminosity distance of
the source, which can be extracted from the amplitude of the inspiral
waveform (assuming the standard cosmological model\footnote{As
  discussed, e.g., in
  Refs.~\cite{EspositoFarese:2003ze,Tattersall:2018map}, theories with
  a dimensionful coupling do not significantly affect the cosmological
  model, and thus the relation between cosmological distance and
  redshift can be assumed to be that predicted by GR.}).
We will consider $p$ as fixed (in modified theories of gravity
  it is typically an integer, or possibly a rational number) so that the number of parameters is the same as in Case~I
  [cf. Eq.~\eqref{param}].
Note that $\alpha$ can be again reabsorbed within $\delta w^{(n)}$ and
$\delta t^{(n)}$, but $\alpha$ is dimensionful if $p\neq0$ (and
so are $\delta w^{(n)}$ and $\delta t^{(n)}$ after the rescaling).

Cases~ I (i.e., $p=0$) and II include some of the best studied
modified theories of gravity:
\begin{itemize}
\item $p=0$: theories with dimensionless couplings in the action
  include certain scalar-tensor theories, Einstein-Aether and
  Ho\v{r}ava gravity (to leading
  order)~\cite{Barausse:2013nwa}.
\item $p=4$: this case includes
  Einstein-scalar-Gauss-Bonnet~\cite{Mignemi:1992nt,Yunes:2011we,Pani:2011gy,Maselli:2015tta,Julie:2019sab}
  and dynamical Chern-Simons
  gravity~\cite{Yunes:2009hc,Ayzenberg:2014aka,Maselli:2017kic}. In
  this case $\gamma_i= \beta^2/(M^{\rm s}_i)^4$, where $\beta$ is the
  coupling constant in the action, which has dimensions of a length
  squared in geometrical units.
\item $p=6$: this case corresponds (e.g.) to certain classes of
  effective field theories~\cite{Cardoso:2018ptl}.
\end{itemize}

As discussed in~\cite{McManus:2019ulj}, if different classes of
gravitational perturbations are non-degenerate at order zero in the
coupling parameter, the leading-order corrections to the QNM
frequencies coming from ${\cal O}(\alpha)$ terms in the action are
${\cal O}(\alpha^2)$. In this case, an operator with mass dimension
$p/2$ in the action will lead to a correction $\alpha^2/M^{p}$ in the
QNMs. In the case of degenerate spectra (e.g., for axial and polar
gravitational perturbations) the leading-order corrections to the QNMs
coming from ${\cal O}(\alpha)$ terms in the action are also
${\cal O}(\alpha)$. The special cases discussed above correspond to
terms in the action with mass dimension $0$, $2$ and $3$, respectively.

\noindent {\bf Case~III: individual charges.}
Since the $\gamma_i$'s appearing in Eq.~\eqref{gammai} depend only on
the fundamental coupling and the masses in the source frame, Cases~ I
and II encompass BHs with secondary hair, but not BHs with primary
hair (corresponding to an extra charge which does not depend on the
mass and spin of the BH).
A simple example of BHs with primary hair are
Kerr-Newman BHs, which are useful and well-studied toy models for
beyond-Kerr BHs, and may be astrophysically significant in certain
dark-sector scenarios~\cite{Cardoso:2016olt}. In the case of BHs with
primary hair we have
\begin{equation}
 \gamma_i=\frac{Q_i^2}{(M^{\rm s}_i)^2}\,, \label{gamma3}
\end{equation}
where $Q_i$ is the charge of the $i$-th source. The number of
parameters necessary to parametrize the spectrum then becomes
\begin{equation}
 {\cal P}'={\cal P}+N=2(D+1) q+N\,.
\end{equation}

Our approach is perturbative by assumption (i.e. $\gamma_i\ll1$). In
the Kerr-Newman example, this means that it can only accommodate
weakly charged BHs.

\subsection{Detection strategies}\label{sec:strategies}
In summary, we can parametrize the QNM spectrum beyond GR with
${\cal P}$ parameters in Cases~I and II, and ${\cal P}'$ parameters in
Case~III. On the other hand, \emph{in principle} for $N$ sources and
$q$ modes we have a certain number ${\cal O}$ of observables, which
depend on whether we consider the ringdown only, or rather extract
$M_i$ and $\chi_i$ from the full inspiral-merger-ringdown waveform
using numerical relativity fits (see e.g.~\cite{Healy:2014yta}) or
analytical models.

\noindent {\bf Using the inspiral-merger-ringdown.}
In this case we measure the individual binary component properties from 
the inspiral-merger-ringdown waveform, and we use numerical-relativity fits 
in GR to evaluate the final masses (in the detector frame) and spins in GR, 
$M_i$ and $\chi_i$ ($i=1,...,N$). This procedure allows us to use 
only the $l=m=2$ QNM to perform BH spectroscopy: it is essentially an extension of inspiral-merger-ringdown consistency 
checks that allows for a non-GR template in the ringdown.

In this case, for $N$ sources and $q$ modes we would have
\begin{equation}
 {\cal O}=2N\times q
\end{equation}
observables, i.e. the frequencies and damping times of $q$ modes. Then
we need
\begin{equation}
 N> \left\{\begin{array}{ll}
            D+1		\qquad\qquad &{\rm Cases~I~and~II,} \\
	    \frac{2 (D+1) q}{2 q-1} 	\qquad\qquad &{\rm Case~III,} 
           \end{array}\right.
\end{equation}
in order to have more observables than parameters. In this case also
$q=1$ is allowed, i.e. the detection of the $l=m=2$ mode for all
sources is sufficient to perform the test. For the minimal case $q=1$
(detection of one mode for each source) we get
\begin{equation}
 N\geq \left\{\begin{array}{ll}
            D+1 		\qquad\qquad &{\rm Cases~I~and~II,} \\
            2(D+1)	 	\qquad\qquad &{\rm Case~III.} 
           \end{array}\right. \label{Nlarger2}
\end{equation}

\noindent {\bf Using the ringdown only.}
For $N$ sources and $q$ modes we have
\begin{equation}
 {\cal O}=2N\times q- 2N = 2N\times (q-1)
\end{equation}
observables, namely the frequencies and damping times of $q$ modes
minus $2N$ observables (the frequencies and damping times of the 
fundamental modes) which have been used to extract the GR masses 
and spin. By comparing ${\cal O}$ with the number of parameters (either
${\cal P}$ or ${\cal P}'$), we need
\begin{equation}
 N> \left\{\begin{array}{ll}
            \frac{q(1+D)}{q-1}		\qquad\qquad &{\rm Cases~I~and~II,} \\
            \frac{2 (D+1) q}{2 q-3} 	\qquad\qquad &{\rm Case~III,} 
           \end{array}\right.
\label{Ocondition}
\end{equation}
in order to have more observables than parameters. Note that this
condition is never satisfied for $q=1$, as expected, since a single
mode can only allow us to determine the masses and spins for each
source.  For the minimal case $q=2$ (i.e., we are detecting $2$ modes
for each source) we get
\begin{equation}
 N\geq \left\{\begin{array}{ll}
            2D+2 		\qquad\qquad &{\rm Cases~I~and~II,} \\
            4D+4	 	\qquad\qquad &{\rm Case~III.} 
           \end{array}\right. \label{Nlarger1}
\end{equation}
Equation~(\ref{Ocondition}) implies $N\geq D+1$ in all cases in the
limit $q\gg1$, i.e. when we can detect several modes for each source.
Note that in this case we need twice of the sources that we needed in the
inspiral-merger-ringdown case [Eq.~(\ref{Nlarger2}) above].

\noindent {\bf Minimum number of sources.}
When the relations~\eqref{Nlarger2} and \eqref{Nlarger1} are saturated, 
$N$ is the minimum number of sources necessary to perform the test, 
whereas further sources will allow for multiple, independent tests. The 
minimum number of sources depends on the truncation order of the 
spin expansion.

\begin{figure}[t]
\centering
\includegraphics[width=0.48\textwidth]{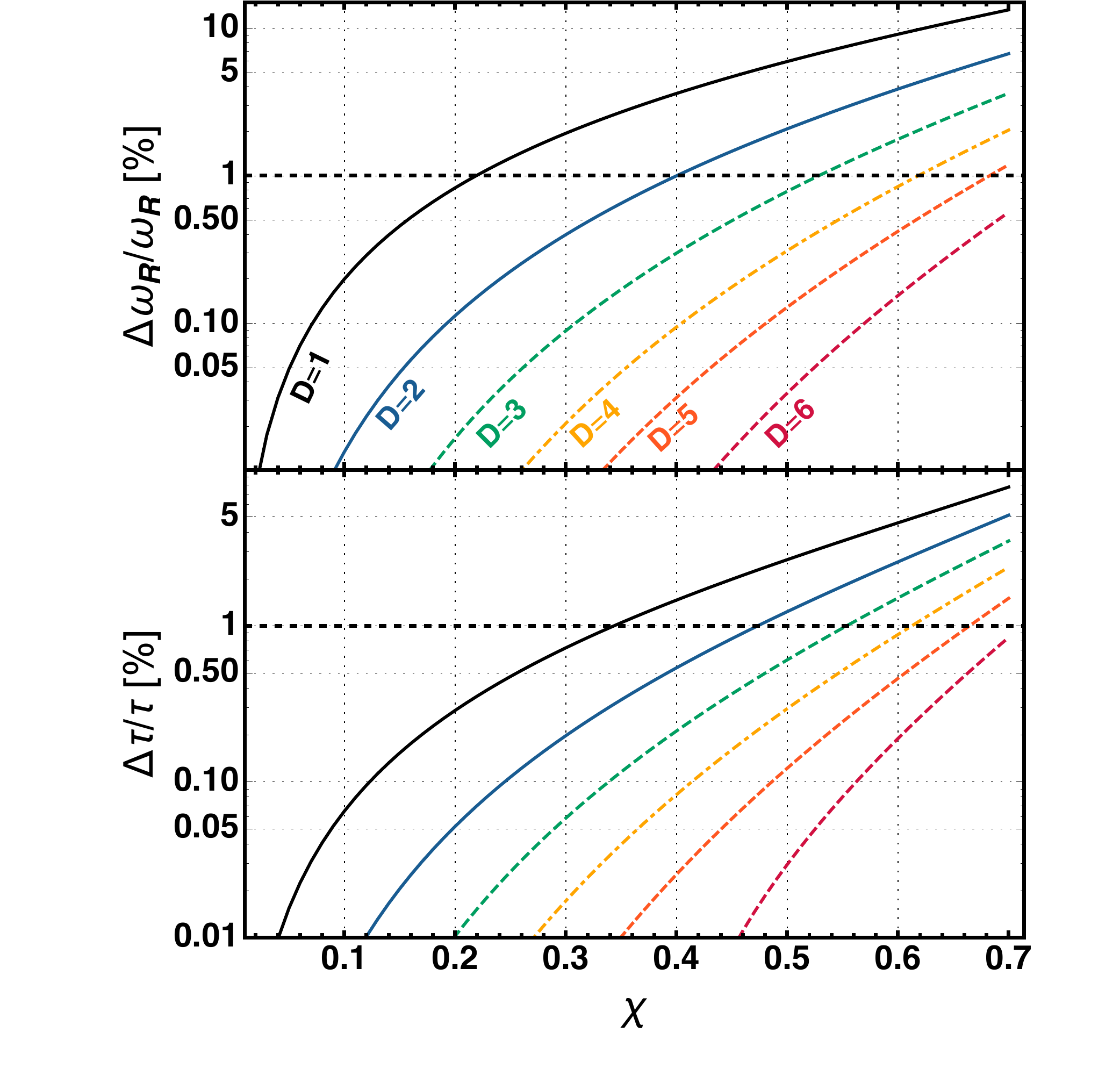}
\caption{Relative percentage errors between the exact $l=m=2$ QNM of 
a Kerr BH (obtained using data from  Ref.~\cite{ringdown}) and their 
small-spin expansion for various truncation orders (see Table~\ref{tab:Kerr}) 
as a function of the BH spin (the value of $D$ is indicated in the legend).}
\label{fig:Kerr} 
\end{figure}

In order to estimate the accuracy needed when we truncate the spin
expansion, in Fig.~\ref{fig:Kerr} we compare numerical calculations of
the QNM frequencies of a Kerr BH with their small-spin expansion at
various truncation orders.
For $D\geq4$ (resp., $D\geq5$) the errors are smaller than $1\%$ for
both the frequency (top panel) and damping time (bottom panel) when
$\chi<0.6$ (resp., $\chi<0.7$).  Therefore a truncation order $D=4$ or
$D=5$ should be sufficient to compute the modes with an accuracy
always better than $1\%$ up to spin $\chi=0.7$. Reaching the same
accuracy at $\chi=0.8$ will require $D\geq7$.

As a proof of principle of the {\sc ParSpec} formalism, in the following we 
shall consider $D=4$ as a working assumption. We shall furthermore restrict 
to the simplest case (Case~I), which minimizes the number of parameters 
and is more in line with existing parametrized tests, e.g. in the 
inspiral~\cite{Will:2014kxa,Yunes:2009ke,Agathos:2013upa}.
Hence, we require $N\geq 5$ inspiral-merger-ringdown detections.
Note that the minimum value of $N$ grows only linearly with $D$: even
a very large spin expansion order (e.g., $D=10$) would require the
same order of magnitude in terms of ringdown detections. Such a number
of detections (even at large SNRs) may well be achievable with LISA
and 3G detectors.

\section{Statistical analysis: Constraining the beyond-Kerr ringdown parameters}
\label{sec:stat}
We use Eqs.~\eqref{model1}-\eqref{model2}, expanded up to fourth order in 
the spin, as templates to interpret the observed frequencies and damping 
times. We assume the {\it true} $(\omega_i,\tau_i)$ to correspond to a Kerr 
BH in GR, i.e. we assume the ``true'' beyond-Kerr parameters $\delta w^{(n)}$ 
and $\delta t^{(n)}$ to be zero.

Our goal is to reconstruct the probability distribution of the beyond-Kerr parameters. 
We consider either $q=1$ or $q=2$, i.e.  either one or 
two modes\footnote{In the $q=2$ case, and for the distributions of binary masses 
and spins discussed below, we have used the excitation factors from 
Refs.~\cite{Baibhav:2017jhs,Baibhav:2018rfk} to check that the second most 
excited mode is $l=m=3$ (rather than $l=2$, $m=1$) roughly $90\%$ of the 
times. To simplify the analysis, we therefore assume that the two measured 
QNMs are the $l=m=2$ and the $l=m=3$ fundamental modes for all sources.} 
detected for each of the $N$ sources.
The purpose of this analysis is to compute the minimum value of the deformation parameters $\delta w^{(n)}$ 
and $\delta t^{(n)}$ which yield a ringdown observation consistent with GR. We 
consider a ground-based 3G detector (ET in the so called ET-D configuration~\cite{Hild:2009ns}) 
and the planned space mission LISA~\cite{Audley:2017drz} as representatives 
of our best near-future chances to carry out BH spectroscopy over a large 
mass range~\cite{Berti:2016lat,Maselli:2017kvl}.

Each source ($i=1,2,...,N$) provides frequencies and damping times 
$(\omega^{(J)}_{i,\,\text{obs}},\tau^{(J)}_{i,\,\text{obs}})$ for $J=1,\ldots, q$ 
modes, with associated parameter estimation errors $\sigma[{\omega_i^{(J)}}]$ 
and $\sigma[{\tau_i^{(J)}}]$ and correlations coefficients.

The values of $\omega^{(J)}_{i,\,\text{obs}}$ and $\tau^{(J)}_{i,\,\text{obs}}$ 
injected in our analysis are computed as follows. We consider the merger 
remnant of $N$ binary coalescences. The $2N$ masses of the binary 
components are drawn from a log-flat distribution between $[5,95]M_\odot$ 
for stellar-origin BHs, and from a uniform distribution within $[10^6,10^7]M_\odot$ 
for massive BHs. For stellar-origin BHs we also require that $m_1+m_2<100M_\odot$ 
\cite{alLIGOScientificCollaborationandVirgoCollaboration:2018cn,Belczynski:2016jno}. 
In both mass ranges the spins are sampled from a uniform distribution 
$\in[-1,1]$.
For illustration, we fix the source distance by choosing the SNR of the first ringdown mode to be $10^2$ for ET and 
$10^3$ for LISA, respectively.  We then compute the mass and the spin of the final 
BH formed after merger using semianalytical relations based on numerical 
relativity simulations in GR~\cite{Healy:2014yta} (as discussed in 
Sec.~\ref{sec:framework}, these are the mass and spin that the final BH 
would have if GR is the correct theory of gravity).
Given the final mass $M_{i}$ and spin $\chi_{i}$ ($i=1\ldots N$), we compute 
the errors on the modes through a Fisher-matrix approach. Following \cite{Berti:2006jo}, 
we roughly estimate the energy in the secondary mode to be $10\%$ of the 
$l=m=2$ mode energy for all sources.
The main purpose of these choices is to test our data-analysis framework, 
and we plan to implement more astrophysically realistic models in future 
work.
  
We use a Bayesian approach to sample the probability distributions 
$\mathbold{P}$ of the model parameters $\vec{\theta}$ for a given set of 
ringdown observations $\vec{d}$. By Bayes' theorem 
$\mathbold{P}(\vec{\theta}|\vec{d})\propto \mathcal{L}(\vec{d}|\vec{\theta}) \mathbold{P}_0(\vec{\theta})$, 
where $\mathcal{L}(\vec{d}|\vec{\theta})$ is the likelihood function and $\mathbold{P}_0(\vec{\theta})$ 
is the prior on the parameters.  For each event, the likelihood function is 
chosen to be Gaussian:
\begin{equation}
 {\cal L}_i(\vec{d}\vert\vec{\theta})={\cal N}(\vec{\mu}_i,\Sigma_i)\,,
\end{equation}
where the vector $\vec{\mu}_i$ depends on the difference between the 
observed $J=1,\ldots q$ modes and the parametrized templates 
\eqref{model1}-\eqref{model2}:
\begin{equation}
\vec{\mu}_i=(\vec{\mu}_i^{(1)},\cdots,\vec{\mu}_i^{(q)})^T\,,
\end{equation}
where each $\vec{\mu}_i^{(J)}$ is a two-component vector
\begin{equation}
\vec{\mu}_i^{(J)}=
\begin{pmatrix}
\omega^{(J)}_{i}-\omega^{(J)}_{i,\,\text{obs}}\\
\tau^{(J)}_{i}-\tau^{(J)}_{i,\,\text{obs}}
\end{pmatrix}\ ,
\end{equation}
and $\Sigma_i$ 
is the covariance matrix that
includes errors and correlations between the frequencies and damping
times measured for the $i$-th source.
Under our assumptions the observed
QNMs correspond to different values of $l$ and $m$, i.e. they are
``quasi-orthonormal'' in the terminology of
Ref.~\cite{Berti:2005ys}. As a consequence the covariance matrix
$\Sigma_i={\rm diag}(\Sigma^{(1)}_i\ldots\Sigma^{(q)}_i)$ is
block-diagonal with each block corresponding to the $J$-th mode, and
the likelihood function can be written as a product of Gaussian
distributions:
\begin{equation}
{\cal N}(\vec{\mu}_i,\Sigma_i)=\prod_{J=1}^q {\cal N}(\vec{\mu}^{(J)}_i,\Sigma^{(J)}_i)\ .
\end{equation}
Moreover, given $N$ independent BH detections, the combined likelihood 
function of the {\sc ParSpec} parameters can be further factorized as
\begin{equation}
{\cal L}(\vec{d}\vert\vec{\theta})=\prod_{i=1}^N{\cal L}_i(\vec{d}\vert\vec{\theta})
=\prod_{i=1}^N\prod_{J=1}^q {\cal N}(\vec{\mu}^{(J)}_i,\Sigma^{(J)}_i)\ .
\end{equation}

The full posterior is obtained through a Markov chain Monte Carlo~(MCMC) 
method based on the Metropolis-Hastings algorithm~\cite{Gilks:1996}, in which 
the proposal matrix is updated through a Gaussian adaptation which enhances 
the convergence to the target distribution~\cite{1085030,5586491}. For each 
dataset we compute $4$ chains of $5\times 10^6$ points, with a thinning factor 
of $0.02$ to reduce the correlation between the samples. We discard $10\%$ 
of the initial points as a burn in.  

The beyond-Kerr dimensionless parameters $\vec{\theta}=\{\delta w_i^{(J)},\delta t_i^{(J)}\}$ 
are sampled by assuming flat distributions within the interval $[-0.5,0.5]$.
Moreover, for simplicity we assume the parameter $p$ associated with 
the mass dimension of the coupling to be $p=0$. We defer a more detailed 
investigation for different values of $p$ to future works. We remark that when 
$p>0$ the dimensionless coupling $\gamma_i =\alpha/M^p$ is much smaller 
for massive BH mergers, and therefore we expect QNM frequency corrections 
for LISA sources to be much smaller than those for stellar-origin BHs.

\begin{figure}[t]
\centering
\includegraphics[width=0.5\textwidth]{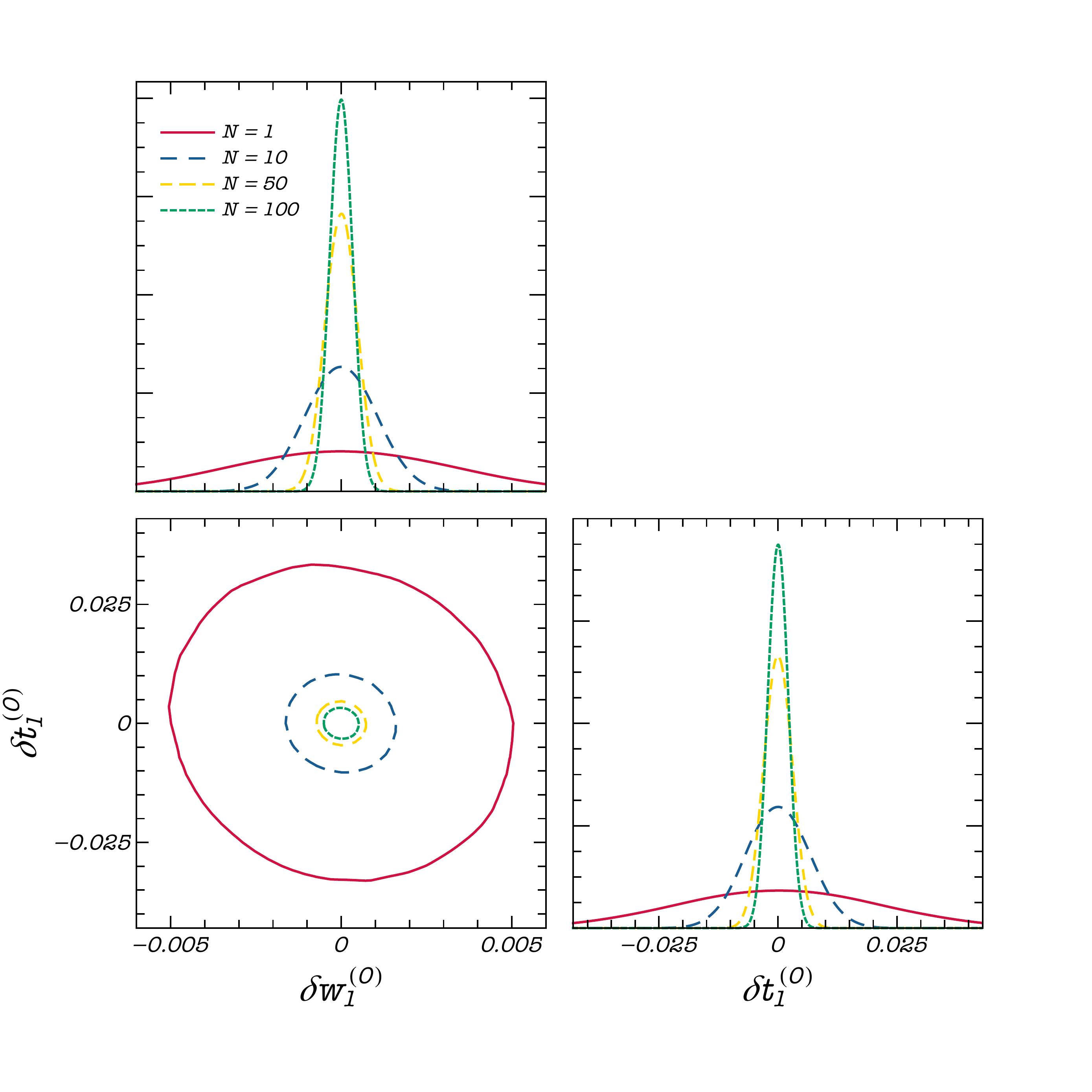}
\caption{Posterior distributions for the beyond-Kerr ringdown
  parameters $\delta w_1^{(0)}$ and $\delta t_1^{(0)}$ inferred
  through the analysis of the $l=m=2$ fundamental QNM for ET,
  assuming $D=0$ and $q=1$. BH masses and spins are estimated 
  from the inspiral-merger-ringdown signal. Colors correspond to 
  different numbers of detections $N$.}
\label{fig:ETposterior} 
\end{figure}

\subsection{Projected constraints with ET}\label{sec:ETanalysis}

\subsubsection{Nonspinning black holes, one mode ($D=0$, $q=1$)}
As a first case study we consider nonrotating BHs, i.e. we assume that 
$D=0$ in the spin expansion of the parametrized templates \eqref{model1} 
and \eqref{model2}. If only the $l=m=2$ mode is detected for each source 
($q=1$), the number of \PS parameters to constrain reduces to 
$\delta w_1^{(0)}$ and $\delta t_1^{(0)}$.
We also focus on the first strategy discussed in Sec.~\ref{sec:strategies}, 
i.e. we assume that the BH masses and spins are measured using the full 
inspiral-merger-ringdown signal. Then the minimum number of events 
required to perform our test is $N_{\rm min}=D+1=1$.

The top and right panels of Fig.~\ref{fig:ETposterior} show the
inferred marginalized distributions of the two \PS parameters as a
function of the number $N$ of stellar mass sources detected by ET. The
posteriors are peaked around zero and, as expected, they become
narrower as $N$ grows.  In the most optimistic case we consider
($N=100$) we find
$\vert\delta w_1^{(0)}\vert \lesssim 5.4\times 10^{-4}$ and
$\vert \delta t_1^{(0)}\vert \lesssim 3.4\times 10^{-3}$ at 90\%
confidence level.
Since the distributions are nearly symmetrical around the peak, we can
define their width as half of the corresponding confidence interval:
$\sigma^i_{90}=1/2(\theta_i^{\rm max}-\theta_i^{\rm min})$, where
$\theta_i^{\rm min}$ and $\theta_i^{\rm max}$ correspond to the values
of the $i$-th parameter such that
\begin{equation}
\int_{\theta_i^{\rm min}}^{\theta_i^{\rm max}}  \mathbold{P}(\theta_i)d\theta_i=0.9
\end{equation}
for the marginalized posterior.
By fitting $\sigma^i_{90}$ as a function of $N$, we find that it scales
like $\sim N^{-1/2}$ to a very good approximation. The contour plots
in the bottom-left panel of Fig.~\ref{fig:ETposterior} show $90\%$
confidence intervals for the 2D joint distribution of the two
parameters, and they show that the parameters are almost completely
uncorrelated.

\subsubsection{Nonspinning black holes, two modes ($D=0$, $q=2$)}
\begin{figure}[t]
\centering
\includegraphics[width=0.4\textwidth]{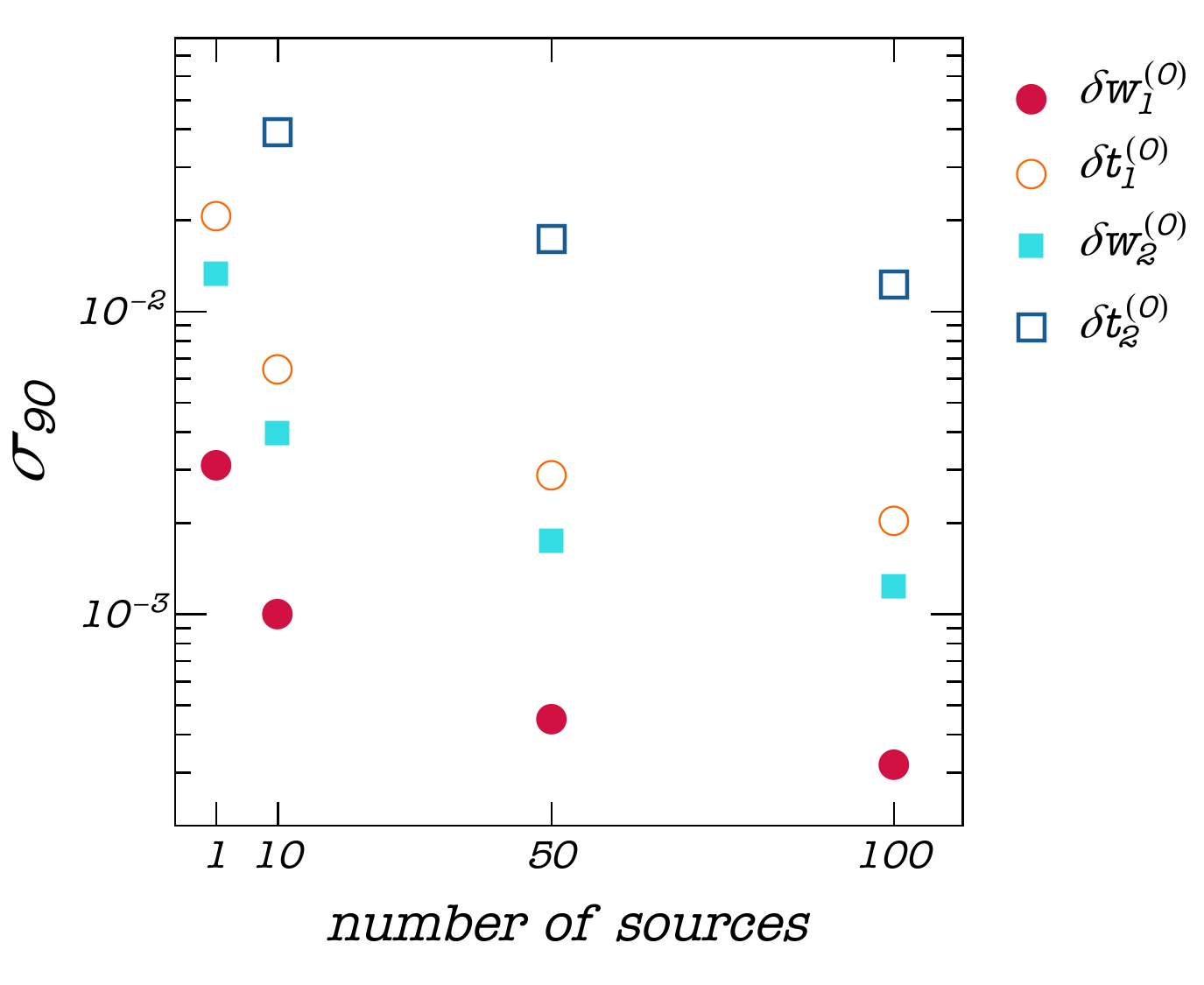}
\caption{90\% confidence intervals $\sigma_{90}$ for the posterior
  distributions of the parameters $\delta w$ and $\delta t$ of the
  first and second mode, as a function of the number of sources
  analysed and considering ET. We consider the case of nonrotating BHs, i.e. $D=0$, assuming that frequencies and 
  damping times are measured through the inspiral-merger-ringdown signal. Only
  measurable parameters are shown.}
\label{fig:ETposterior2} 
\end{figure}

Our approach can accommodate an arbitrary number of modes. As a
slightly more complex scenario, we still set $D=0$ but we now consider
the observation of the primary ($l=m=2$) and secondary ($l=m=3$) QNM
for each BH (i.e., $q=2$), thus doubling the number of parameters that
we wish to constrain. Figure~\ref{fig:ETposterior2} shows the width
$\sigma$ of the sampled posteriors as a function of $N$. The smallest
values of $\sigma$ (i.e., the strongest bounds) correspond to the
frequency corrections $\delta w^{(0)}_1$ and $\delta w^{(0)}_2$ of the
primary and secondary mode, respectively. The widths are larger for
corrections to the damping times ($\delta t^{(0)}_1$ and
$\delta t^{(0)}_2$), which are typically harder to measure. Most
importantly, Figure~\ref{fig:ETposterior2} shows that some of the \PS
parameters can become measurable by increasing the number of
observations $N$: for example, for $N=1$ the marginalized distribution
of $\delta t_2^{(0)}$ is flat within the allowed range of values, and
hence unconstrained by the data, but this quantity can be constrained
for larger values of $N$. As in the $q=1$ case considered
  above, we find that the width of all parameters scales as
  $\sim N^{-1/2}$ to a very good approximation when the number of
  sources is large enough (typically of the order of $N\sim100$).

\begin{figure}[t]
\centering
\includegraphics[width=0.4\textwidth]{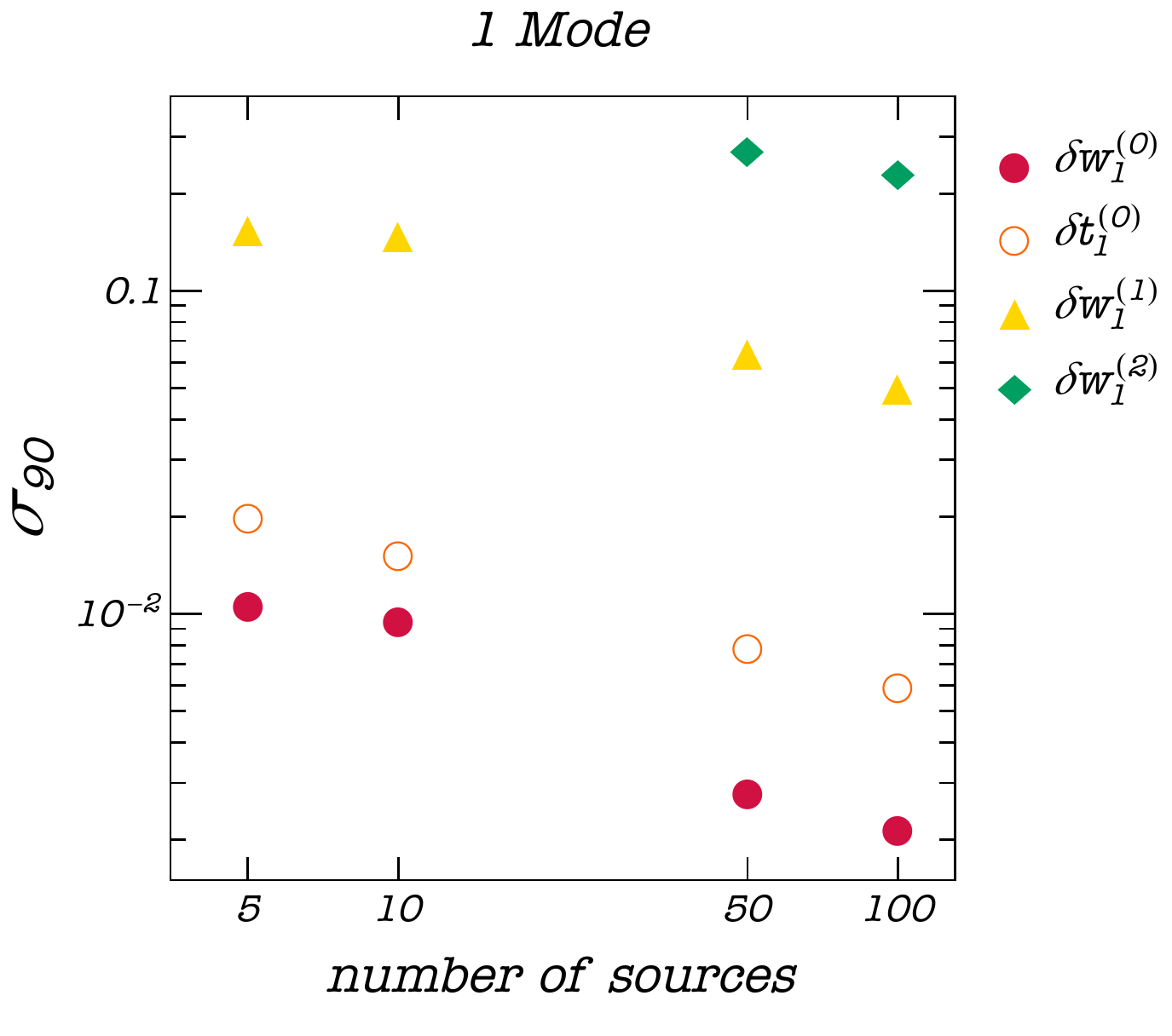}\vspace{0.5cm}
\includegraphics[width=0.41\textwidth]{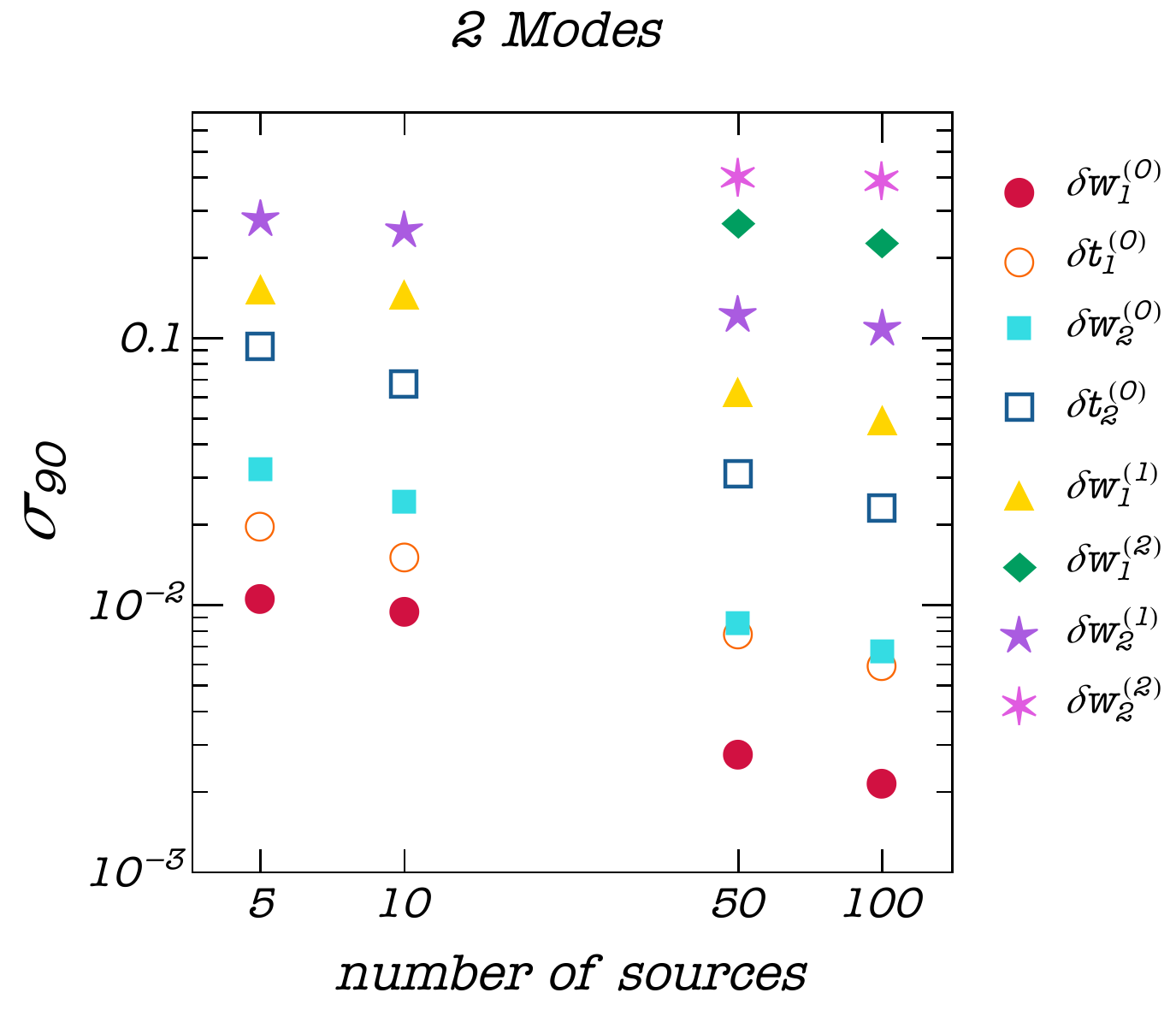}
\caption{90\% confidence intervals $\sigma_{90}$ of the beyond-Kerr
  ringdown spin coefficients considering GR modifications up to fourth
  order in rotation ($D=4$) as a function of the number of sources $N$
  observed by ET. Top and bottom panels refer to $q=1$ and $q=2$,
  respectively. Only measurable parameters are shown.}
\label{fig:ETposterior3} 
\end{figure}

\begin{figure}[!htbp]
\centering
\includegraphics[width=0.4\textwidth]{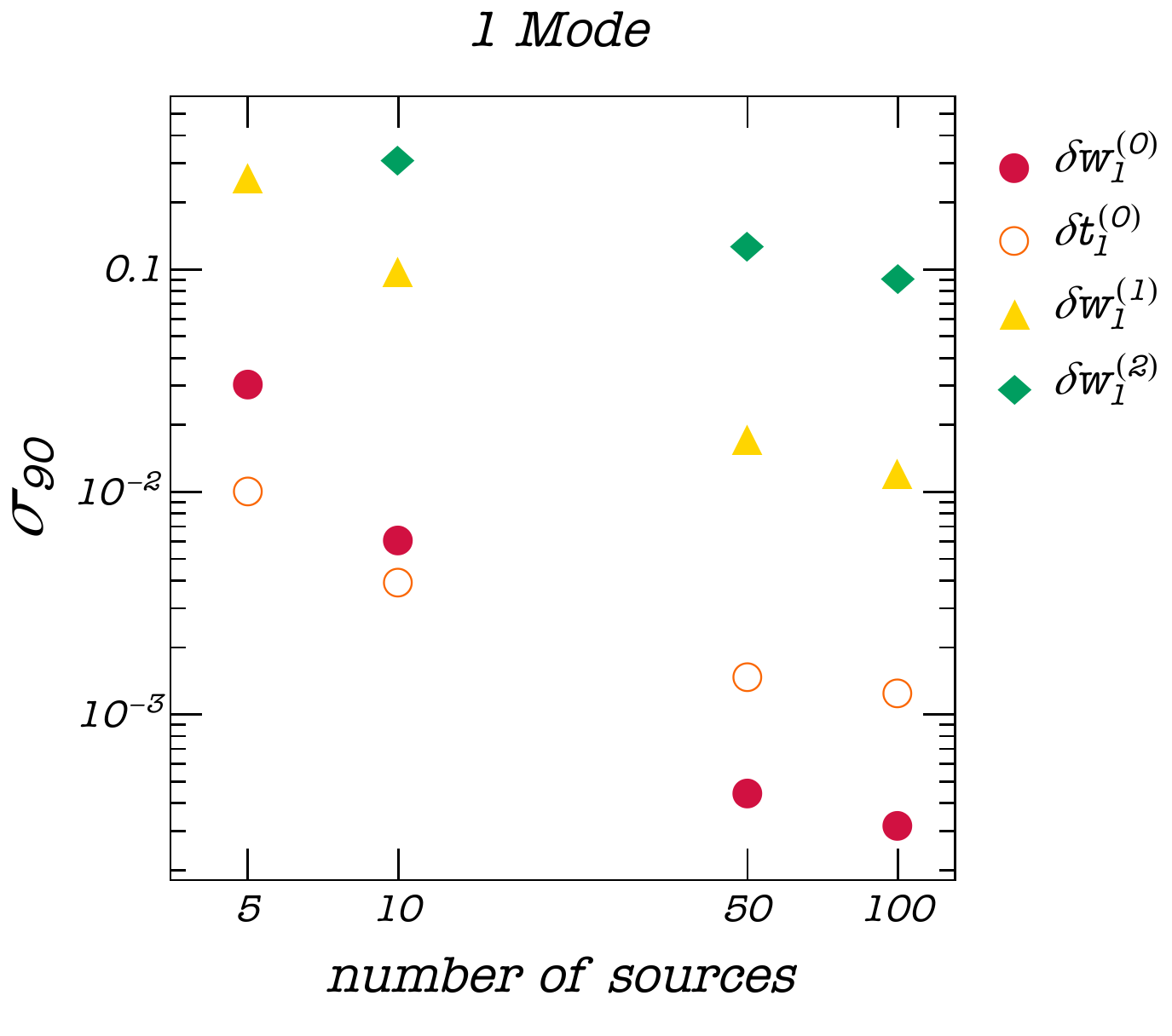}\vspace{0.5cm}
\includegraphics[width=0.4\textwidth]{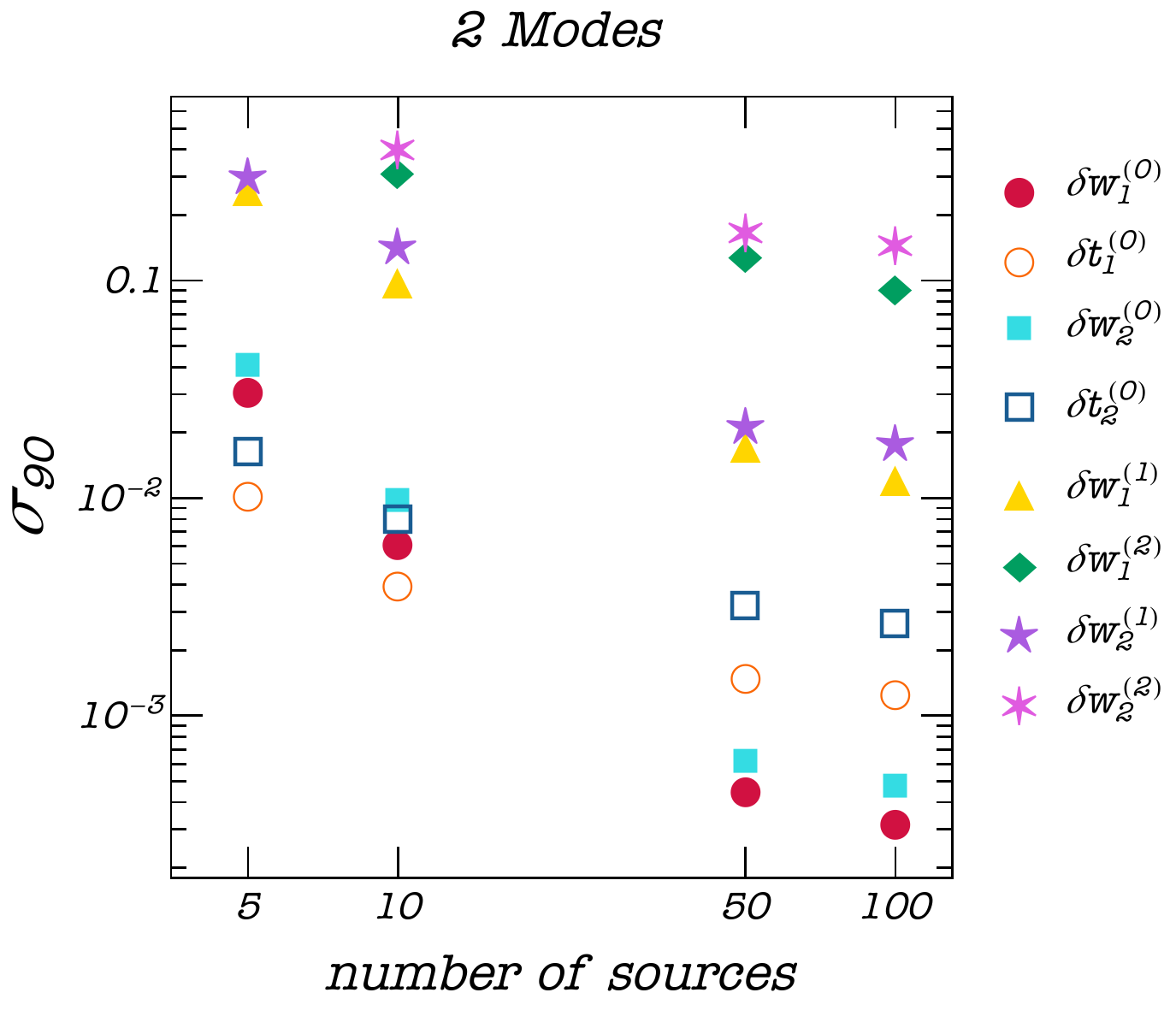}
\caption{Same as Fig.~\ref{fig:ETposterior3}, but for massive BHs observed by LISA.}
\label{fig:LISA} 
\end{figure}

\begin{figure}[!htbp]
\centering
\includegraphics[width=0.45\textwidth]{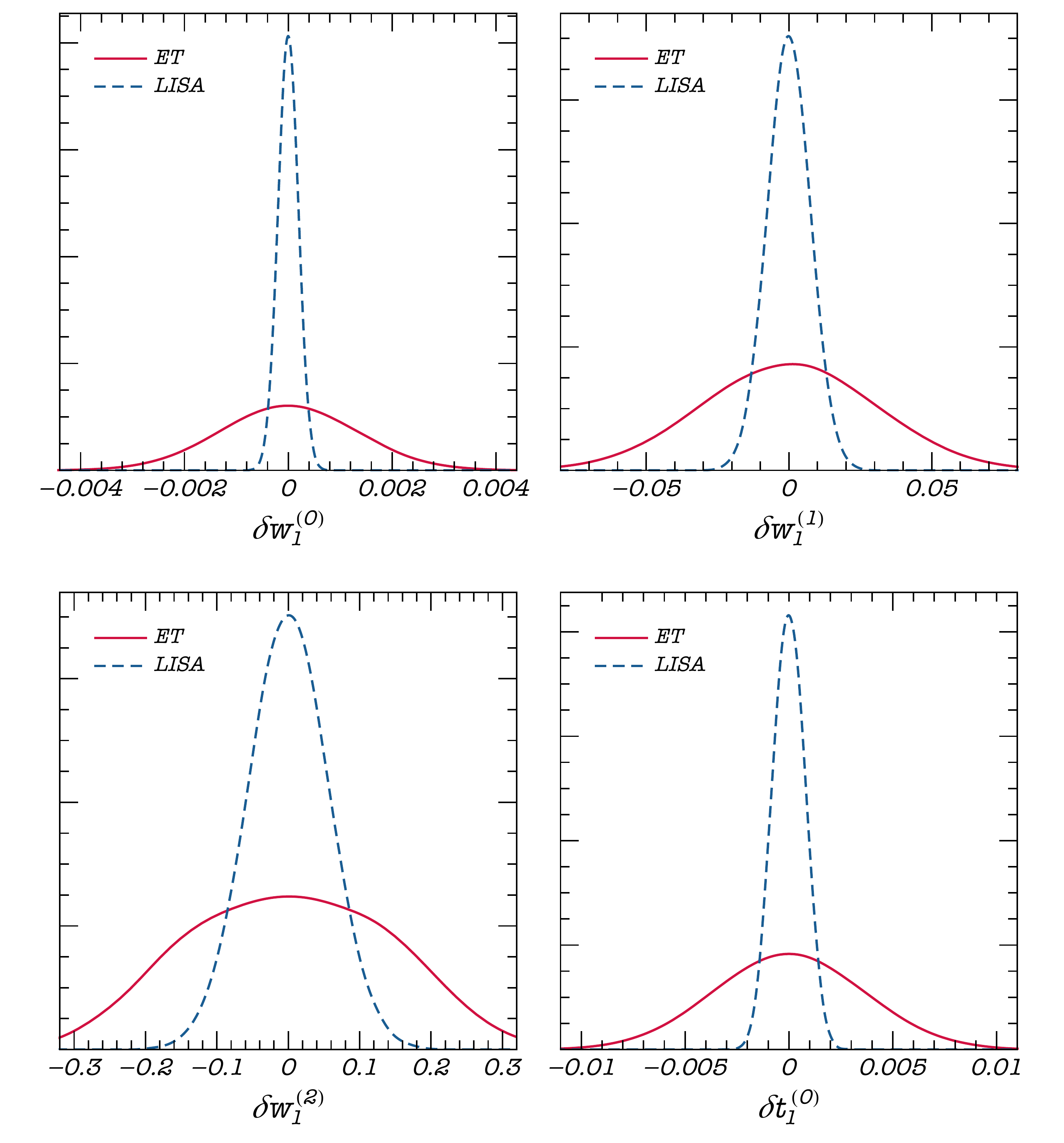}
\caption{Comparison between the posteriors of some of the \PS 
parameters obtained for ET and LISA assuming spinning corrections with 
$D=4$ for $N=100$ sources. The probability distribution refers to 
the same dataset shown in the top panel of Fig.~\ref{fig:LISA}.}
\label{fig:comparison} 
\end{figure}

\subsubsection{Spinning black holes ($D=4$, $q=1$ and $q=2$)} 
We can turn to the more realistic scenario of spinning BHs. We truncate the 
expansion at $D=4$, so that modes in GR are estimated with an accuracy 
better than $1\%$ up to spins $\chi\sim 0.6$ (see Fig.~\ref{fig:Kerr}). Results 
for ET observations with $q=1$ and $q=2$ are shown in the top and bottom 
panels of Fig.~\ref{fig:ETposterior3}, respectively.

Consider first $q=1$ (top panel). Even with a large number of detections $N$, 
only some of the parameters are measurable. In general we can constrain 
with good accuracy the first three $\delta w_1^{(n)}$'s, i.e. GR deviations up 
to quadratic order in the spin ($n=0,\,1,\,2$). By contrast, only the non-spinning 
correction $\delta t_1^{(0)}$ to the damping times is bounded by the data.

Note however that $\delta w_1^{(2)}$ becomes measurable only for $N\gtrsim 50$. 
Moreover the posterior of the coefficients proportional to the BH's angular 
momentum are more than one order of magnitude wider than the posterior of their nonspinning 
counterparts. For completeness, in Appendix~\ref{app:corner} we show the 
corner plot for the marginalized and joint posterior distributions of the measurable 
\PS parameters for $q=1$. The spin-dependent frequency corrections 
show a correlation which generally decreases with $N$, while $\delta t_1^{(0)}$ 
is typically uncorrelated with the other coefficients.

The $q=2$ case (bottom panel of Fig.~\ref{fig:ETposterior3}) is very similar: the 
width of the posteriors inferred through the MCMC decreases with $N$. The 
hierarchy among the beyond-Kerr parameters is also the same as in the single-mode 
case: zero-order (nonspinning) terms are best constrained, followed by corrections 
that are of low order in rotation. Remarkably, with $N=100$ sources we can put 
tight upper bounds on the coefficient of the secondary mode, with 
$\vert\delta w_2^{(0)}\vert \lesssim 7\times 10^{-3}$
$\vert\delta w_2^{(1)}\vert \lesssim 10^{-1}$ and
$\vert \delta t_2^{(0)}\vert \lesssim 2\times 10^{-2}$.

These results can be straightforwardly adapted to the second detection strategy 
outlined in Sec.~\ref{sec:strategies}. In this case we assume that for each 
observation we extract two QNMs from the postmerger GW signal, using the 
frequency and damping time of the fundamental mode to determine the mass 
and the spin of the source. This scenario is comparable to the single-mode case 
described above, but now we inject into the MCMC the subdominant QNM, which 
has lower SNR. Therefore we expect that the bounds shown in Fig.~\ref{fig:ETposterior2} 
and in the top panel of Fig.~\ref{fig:ETposterior3} would worsen by roughly one 
order of magnitude.

\subsection{Projected constraints with LISA}
 
We now perform a similar Bayesian analysis for LISA QNM observations
of massive BH binary mergers. For nonrotating BHs ($D=0$) the
reconstructed posteriors are nearly a factor of ten tighter than the
corresponding distributions for ET. This is somehow expected, since
the parameters are nearly uncorrelated and the inference is dominated
by the SNR of the detections (which we assumed to be one order of
magnitude larger for LISA than for ET).

Figure~\ref{fig:LISA} shows the width of the posteriors for spinning
BHs with spin corrections up to $D=4$ and $q=1$ (top panel) or $q=2$
(bottom panel). The values of the upper bounds on the \PS parameters
are in qualitative agreement with those obtained for ET in
Fig.~\ref{fig:ETposterior3}.

To facilitate comparisons, in Fig.~\ref{fig:comparison} we show the
posterior distributions inferred from a sample of $N=100$ observations
with ET and LISA, assuming $q=1$. At least in the case of scale-free
corrections considered in this work ($p=0$), LISA constraints are more
stringent.  For the best constrained parameters, the $90\%$ confidence
intervals with LISA are
$\vert\delta w_1^{(0)}\vert\lesssim 3.2\times 10^{-4}$
($\vert\delta t_1^{(0)}\vert\lesssim 1.3\times 10^{-3}$), which are
smaller than the corresponding values obtained for ET by a factor
$\sim6$ ($\sim5$). For the upper bounds on the spinning coefficients
we get $\vert\delta w_1^{(1)}\vert\lesssim 1.2\times 10^{-2}$ and
$\vert\delta w_1^{(2)}\vert\lesssim 9.3\times 10^{-2}$ which are
$\sim4$ and $\sim3$ times smaller than those inferred by ET.  The
corner plot in Fig.~\ref{fig:LISAcorner} of Appendix~\ref{app:corner}
shows the complete set of marginalized and joint distributions derived
for such parameters. As already discussed in
Sec.~\ref{sec:ETanalysis}, the coefficients $\delta w_i^{({J)}}$ that
modify the mode's frequencies are all correlated to each other, while
the correction to the damping time is almost decoupled from the other
parameters.

The two-mode analysis ($q=2$) follows the same trend. Moreover,
comparing the bottom panels of Figs.~\ref{fig:ETposterior3} and
\ref{fig:LISA} we note that --~unlike ET~-- LISA will be able to
constrain possible deviations from the primary and secondary modes
with comparable accuracy.

\section{Possible extensions}
\label{sec:future}
In this work we have presented a data-analysis framework (that we dub \PS) and performed
a preliminary analysis. Here we discuss several interesting extensions
that should be explored in the future.

In our proof-of-principle data-analysis demonstration we consider only
scale-free corrections, i.e. $p=0$. The extension to different values
of $p$ (and hence to dimensionful couplings) is technically
straightforward, but introducing a scale inevitably makes certain
sources more relevant than others. Specifically, for a coupling
parameter $\alpha$ with mass dimension $p$, sources with
(source-frame) mass $M^s$ such that $\alpha/(M^s)^p\sim 0.1$ will
contribute the most, whereas sources with $\alpha/(M^s)^p\ll1$ will be
irrelevant for the analysis. We could simply consider only the subset of
events such that $\alpha/(M^s)^p$ is larger than a fixed threshold.
Overall, this would require more detections.
 
The assumption that our ``true'' signal is the standard ringdown
within GR allows us to put at most upper bounds on the beyond-Kerr
ringdown parameters, but we can \emph{search} directly for GR
deviations by writing the beyond-Kerr ringdown parameters explicitly
for a given theory.  As an extension, it would be interesting to
consider a particular non-GR theory and to recover the ringdown signal
in this theory with a standard GR ringdown template, in order to
quantify systematic errors~\cite{Vallisneri:2013rc}.
  
Additional ``branches'' of the QNM spectrum which are not
perturbatively close to the Kerr spectrum are expected in virtually
any extension of GR, although they might be excited with small
amplitude (see
e.g.~\cite{Molina:2010fb,Blazquez-Salcedo:2016enn,Tattersall:2017erk,Okounkova:2017yby,Witek:2018dmd,Okounkova:2019dfo}). To
leading order, the extra modes coincide with the corresponding QNMs of
a Kerr BH in GR: for example, extra scalar~(vector) degrees of freedom
can give rise to standard scalar~(vector) QNMs in the gravitational
waveform, with amplitude proportional to the coupling parameter of the
theory~\cite{Molina:2010fb,Blazquez-Salcedo:2016enn,Tattersall:2017erk,Witek:2018dmd}. Our
formalism can accommodate extra QNMs, which can be parametrized with
Eqs.~\eqref{model1}--\eqref{model2} with $\gamma_i=0$ by setting
$w^{(n)}$ and $t^{(n)}$ to match the corresponding values for the
(scalar, vector, etcetera) QNMs of a Kerr BH in GR (but
see~\cite{McManus:2019ulj} for possible complications arising when
different perturbations are coupled to each other).
   
Some theories of gravity may have multiple coupling constants, rather
than the single perturbative parameter considered here. It is
straightforward to extend our formalism to this case.

Our spin expansion is necessary to parametrize the ringdown in terms
of a set of constant coefficients (as opposed to {\em functions} of
the spin). The resulting systematic errors can be reduced by
considering higher-order expansions than the $D=4$ case considered
here. To check the impact of the truncation order, we have also
considered a spin expansion truncated at $D=6$. In our tests, the
posterior distribution of the measurable parameters did not change
significantly relative to the $D=4$ case.

An obvious and important extension of our work is to compute rates for
both 3G and LISA sources using more realistic astrophysical models. In
this preliminary analysis we have assumed $10-100$ events at
${\rm SNR}\sim 100$ ($\sim 1000$) for ET (LISA), corresponding to nearby
sources.
It is important to estimate whether these estimates are realistic.
Since we select only large-SNR sources, which are generally the
closest ones, we have neglected the source redshift. Even the closest
LISA ringdown sources may have nonnegligible redshift, and therefore
cosmological effects should be included in a more refined analysis.

We assume that two different angular modes are detected for each
sources. The extension to multiple angular modes is straightforward,
and in general it should lead to stronger constraints. Likewise, in
realistic scenarios not only the amplitude of the secondary mode, but
also its nature will depend on the binary parameters (mass ratio and
spins): in general the second most exited mode will correspond to
$l=m=3$ only for a subset of sources, whereas the mode with $l=2$,
$m=1$ and $l=m=4$ may be dominant for others. Future work should
extend our parameter estimation strategy to the case of multiple,
source-dependent secondary modes.

For any given $(l,\,m)$ we consider only the \emph{fundamental} mode,
neglecting the overtones.  In general, overtones are relevant for
parameter
estimation~\cite{Leaver:1986gd,Berti:2006wq,London:2014cma,Baibhav:2017jhs,Giesler:2019uxc,Isi:2019aib}. However,
the frequencies of different overtones are very closely spaced and
hard to resolve~\cite{Berti:2005ys,Bhagwat:2019dtm}, and therefore
--~besides consistency tests with mass/spin inferred from the whole
waveform~\cite{Giesler:2019uxc,Isi:2019aib}~-- it is hard to use them
for direct BH spectroscopy. If the ringdown SNR is very high (as
expected for 3G detectors and LISA)~\cite{Bhagwat:2019dtm} the
overtones may be resolved, and therefore they should be included in
our model.

\section{Conclusions}
\label{sec:conclusions}

Ringdown tests and BH spectroscopy will allow us to place much tighter
constraints on strong-field gravity when high-SNR BH merger detections
will become routine, as expected for LISA and 3G interferometers. We
have introduced an approach based on ``Parametrized Ringdown Spin
Expansion Coefficients''~({\sc ParSpec}) to parametrize beyond-GR
deviations from the standard QNM ringdown of a Kerr BH in Einstein's
theory. We demonstrated that this method can be used to constrain a
large number of beyond-Kerr ringdown parameters using multiple
ringdown observations.

Our main results can be summarized as follows:
\begin{itemize}
\item At variance with previous frameworks
  (e.g.,~\cite{Meidam:2014jpa}), in {\sc ParSpec} the ringdown
  parameters can be mapped to virtually any (perturbative) extension
  to GR. The framework is perturbative in the spin but can be made
  arbitrarily precise --~at least in principle~-- through high-order
  spin expansions.
\item We estimate that for a spin expansion of order five or higher
  ($D\geq 5$), truncation errors are below $1\%$ for spins
  $\chi\lesssim 0.7$ (see Fig.~\ref{fig:Kerr}).
\item The number of beyond-GR parameters can be very large (especially
  in the case of a high-order spin expansions), but we can use
  Bayesian inference to identify the most easily measurable expansion
  coefficients. It turns out that ${\cal O}(10)$ ringdown detections
  at ${\rm SNR}\sim 100$ (as achievable with ET and Cosmic Explorer)
  can constrain the beyond-Kerr parameters associated to zeroth- and
  first-order corrections in the spin, whereas constraining the
  second-order in spin coefficients will require ${\cal O}(10)$
  ringdown detections at ${\rm SNR}\sim 1000$, something that could be
  achievable with LISA.
\item An important consequence of this observation is that, even
  including beyond-Kerr parameters up to $D=4$ in the spin, only those
  with $D\leq 2$ can be actually measured in the foreseeable future.
\item The method can automatically accommodate an arbitrary number of
  sources. As expected, the posterior distribution becomes narrower as
  the number of events $N$ increases. Their width scales approximately
  as $\sigma\sim N^{-1/2}$ when $N\gtrsim100$ (the accuracy of this
  scaling improves when the number of parameters is not too
  large). Interestingly, the number of beyond-Kerr parameters that can
  be measured increases with $N$. Furthermore, as the number of
  sources increases it could be possible to perform multiple and
  independent checks.
\end{itemize}

Future work will focus on a more systematic analysis of {\sc ParSpec} along the lines discussed in 
Sec.~\ref{sec:future}.

\begin{acknowledgments}

P.P. acknowledges financial support provided under the European Union's 
H2020 ERC, Starting Grant agreement no.~DarkGRA--757480.
This project has received funding from the European Union's Horizon 2020 
research and innovation programme under the Marie Sklodowska-Curie grant 
agreement No 690904.
The authors would like to acknowledge networking support by the COST 
Action CA16104 and support from the Amaldi Research Center funded by 
the MIUR program "Dipartimento di Eccellenza" (CUP: B81I18001170001). 
This project has also received funding from MIUR under the PRIN programme
.B. is supported by NSF Grant No. PHY-1912550, NSF Grant No. AST-1841358, 
NSF-XSEDE Grant No. PHY-090003, NASA ATP Grant No. 17-ATP17-0225, and 
NASA ATP Grant No. 19-ATP19-0051. Computational work was performed at the 
Maryland Advanced Research Computing Center (MARCC).
\end{acknowledgments}

\appendix

\section{QNM parametrization}\label{app:framework}
In this appendix we show that --~for theories which are perturbatively close to GR~-- the QNM frequencies and damping 
times are given by Eqs.~\eqref{model1} and \eqref{model2}, where $M_i$ and $\chi_i$ are the mass (in the detector frame)
and spin of the 
$i$-th source, \emph{both measured assuming GR}. 

\begin{figure*}[!htbp]
\centering
\includegraphics[width=0.8\textwidth]{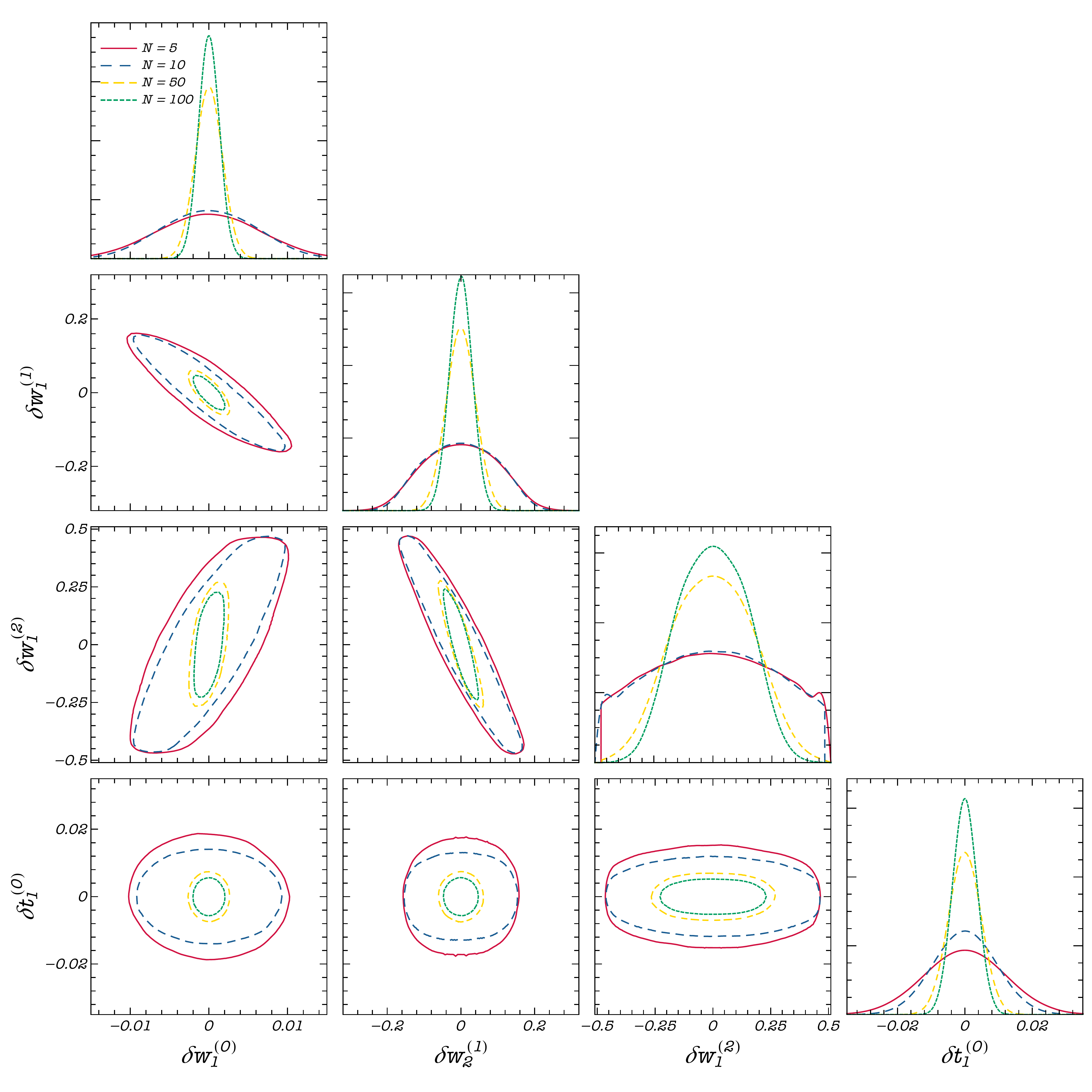}
\caption{Corner plot for the {\sc ParSpec} parameters inferred by ET
  observations of a single QNM from rotating BHs. We consider spinning
  GR corrections up to the fourth order. Diagonal and off-diagonal
  panels show marginalized and joint 2D distributions,
  respectively. Contour plots identify 90\% credible intervals. Only
  measurable parameters are shown in the plot, while the remaining
  $\delta w_J^{(n)}$ and $\delta t_J^{(n)}$ are unconstrained by the
  data.}
\label{fig:ETcorner} 
\end{figure*}

In general, the QNM parametrization can be written similarly to Eqs.~\eqref{model1} and \eqref{model2}, but in terms 
of the \emph{physical} masses and spin and of new parameters, namely
\begin{eqnarray}
 \omega_i^{(J)} &=&\frac{1}{\bar M_i} \sum_{n=0}^D \bar \chi^n_i w^{(n)}_J\left(1+\gamma_i\delta 
W^{(n)}_J\right) \,,\label{model1bis}\\
 \tau_i^{(J)} &=&\bar M_i \sum_{n=0}^D \bar\chi^n_i t^{(n)}_J\left(1+\gamma_i\delta T^{(n)}_J\right) 
\,,\label{model2bis}
\end{eqnarray}
where $\bar M_i$ and $\bar \chi_i$ are the \emph{physical} mass (in the detector frame) and spin of 
the $i$-th source, whereas  $\gamma_i=\alpha(1+z_i)^p/{\bar M}_i^p$.
The physical masses and spins can be expanded as
\begin{eqnarray}
 \bar M_i    &=& M_i(1+\gamma_i \delta m)\,, \\
 \bar \chi_i &=& \chi_i(1+\gamma_i \delta \chi)\,,  
\end{eqnarray}
where $M_i$ and $\chi_i$ are the values \emph{within GR}, whereas $\delta m$ and $\delta \chi$ are universal, 
\emph{dimensionless} corrections due to the 
fact that the underlying theory is not GR. Like $\delta w_J^{(n)}$ and $\delta t_J^{(n)}$ (or, equivalently, $\delta 
W_J^{(n)}$ and $\delta T_J^{(n)}$), these corrections depend only on the theory and not on the source. 

It is easy to check that, to leading order in $\alpha$, the redefinitions
\begin{eqnarray}
 \delta W_J^{(n)} &=& \delta w_J^{(n)}+\delta m - n \delta \chi \,,\\
\delta T_J^{(n)} &=& \delta t_J^{(n)}-\delta m - n \delta \chi \,,
\end{eqnarray}
bring Eqs.~\eqref{model1bis} and \eqref{model2bis} to the form in Eqs.~\eqref{model1} and \eqref{model2} used in the 
main text.

The above redefinitions also show that there is some degeneracy among the beyond-GR parameters, and that one can only 
constrain the combinations $\delta w_J^{(n)}$ and $\delta t_J^{(n)}$, which contains the \emph{intrinsic} mode 
corrections ($\delta W_J^{(n)}$ and $\delta T_J^{(n)}$) and the mass and spin corrections ($\delta m$ and $\delta 
\chi$).

\section{Corner plots for single-mode analysis of the \PS parameters}\label{app:corner}

\begin{figure*}[!htbp]
\centering
\includegraphics[width=0.82\textwidth]{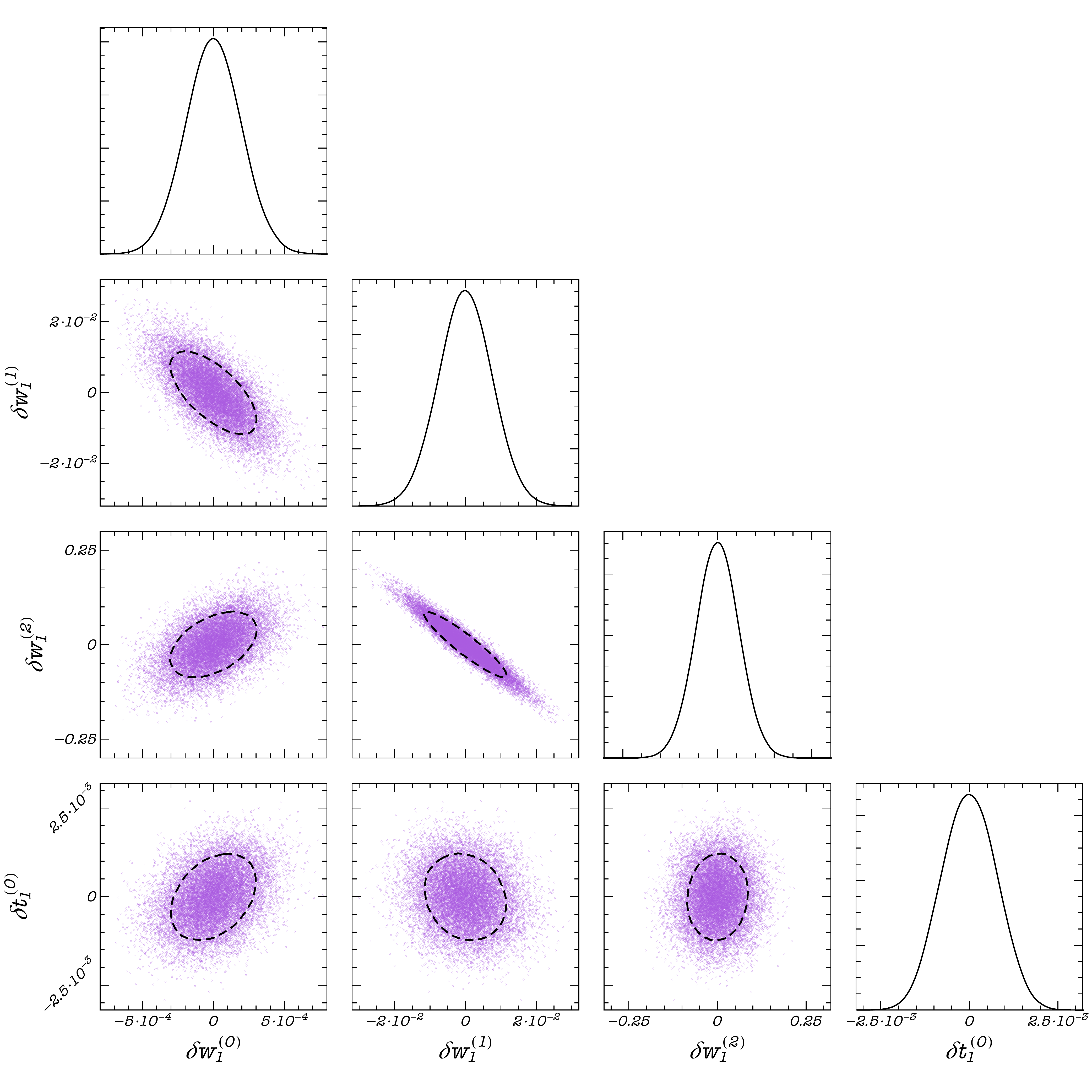}
\caption{Same as Fig.~\ref{fig:ETcorner} but for the parameters
  inferred by LISA using a single QNM, assuming GR corrections up
  to the fourth order, and $N=100$ sources. Contour plots identify 90\% credible intervals,
  while purple dots correspond to MCMC samples.}
\label{fig:LISAcorner} 
\end{figure*}

In Figs.~\ref{fig:ETcorner} and \ref{fig:LISAcorner} we show two corner
plots for the marginalized and joint posterior distributions of the \PS
parameters for the case of ET and LISA detections, respectively. The spin dependent terms feature a correlation which
decreases with the overall number of sampled events. Note that $\delta t_1^{(0)}$ is decoupled from the rest of the 
coefficients.

\bibliographystyle{utphys}
\bibliography{Ref}

\providecommand{\href}[2]{#2}\begingroup\raggedright\begin{thebibliography}{100}

\bibitem{LambShift}
W.~E. Lamb and R.~C. Retherford, ``Fine structure of the hydrogen atom by a
  microwave method,'' \href{http://dx.doi.org/10.1103/PhysRev.72.241}{{\em
  Phys. Rev.} {\bfseries 72} (Aug, 1947) 241--243}.
  \url{https://link.aps.org/doi/10.1103/PhysRev.72.241}.

\bibitem{1980ApJ...239..292D}
S.~{Detweiler}, ``{Black holes and gravitational waves. III - The resonant
  frequencies of rotating holes},''
  \href{http://dx.doi.org/10.1086/158109}{{\em \apj} {\bfseries 239} (July,
  1980) 292--295}.

\bibitem{Dreyer:2003bv}
O.~Dreyer, B.~J. Kelly, B.~Krishnan, L.~S. Finn, D.~Garrison, and
  R.~Lopez-Aleman, ``{Black hole spectroscopy: Testing general relativity
  through gravitational wave observations},''
  \href{http://dx.doi.org/10.1088/0264-9381/21/4/003}{{\em Class. Quant. Grav.}
  {\bfseries 21} (2004) 787--804},
\href{http://arxiv.org/abs/gr-qc/0309007}{{\ttfamily arXiv:gr-qc/0309007
  [gr-qc]}}.

\bibitem{Berti:2005ys}
E.~Berti, V.~Cardoso, and C.~M. Will, ``{On gravitational-wave spectroscopy of
  massive black holes with the space interferometer LISA},''
  \href{http://dx.doi.org/10.1103/PhysRevD.73.064030}{{\em Phys. Rev.}
  {\bfseries D73} (2006) 064030},
\href{http://arxiv.org/abs/gr-qc/0512160}{{\ttfamily arXiv:gr-qc/0512160
  [gr-qc]}}.

\bibitem{Vishveshwara:1970cc}
C.~V. Vishveshwara, ``{Stability of the schwarzschild metric},''
\href{http://dx.doi.org/10.1103/PhysRevD.1.2870}{{\em Phys. Rev.} {\bfseries
  D1} (1970) 2870--2879}.

\bibitem{Chandra}
S.~Chandrasekhar, {\em The Mathematical Theory of Black Holes}.
\newblock Oxford University Press, New York, 1983.

\bibitem{Kokkotas:1999bd}
K.~D. Kokkotas and B.~G. Schmidt, ``{Quasinormal modes of stars and black
  holes},'' \href{http://dx.doi.org/10.12942/lrr-1999-2}{{\em Living Rev. Rel.}
  {\bfseries 2} (1999) 2},
\href{http://arxiv.org/abs/gr-qc/9909058}{{\ttfamily arXiv:gr-qc/9909058
  [gr-qc]}}.

\bibitem{Ferrari:2007dd}
V.~Ferrari and L.~Gualtieri, ``{Quasi-Normal Modes and Gravitational Wave
  Astronomy},'' \href{http://dx.doi.org/10.1007/s10714-007-0585-1}{{\em Gen.
  Rel. Grav.} {\bfseries 40} (2008) 945--970},
\href{http://arxiv.org/abs/0709.0657}{{\ttfamily arXiv:0709.0657 [gr-qc]}}.

\bibitem{Berti:2009kk}
E.~Berti, V.~Cardoso, and A.~O. Starinets, ``{Quasinormal modes of black holes
  and black branes},''
  \href{http://dx.doi.org/10.1088/0264-9381/26/16/163001}{{\em Class. Quantum
  Grav.} {\bfseries 26} (2009) 163001},
\href{http://arxiv.org/abs/0905.2975}{{\ttfamily arXiv:0905.2975 [gr-qc]}}.

\bibitem{Konoplya:2011qq}
R.~A. Konoplya and A.~Zhidenko, ``{Quasinormal modes of black holes: From
  astrophysics to string theory},''
  \href{http://dx.doi.org/10.1103/RevModPhys.83.793}{{\em Rev. Mod. Phys.}
  {\bfseries 83} (2011) 793--836},
\href{http://arxiv.org/abs/1102.4014}{{\ttfamily arXiv:1102.4014 [gr-qc]}}.

\bibitem{Berti:2015itd}
E.~Berti {\em et~al.}, ``{Testing General Relativity with Present and Future
  Astrophysical Observations},''
  \href{http://dx.doi.org/10.1088/0264-9381/32/24/243001}{{\em Class. Quant.
  Grav.} {\bfseries 32} (2015) 243001},
\href{http://arxiv.org/abs/1501.07274}{{\ttfamily arXiv:1501.07274 [gr-qc]}}.

\bibitem{Barack:2018yly}
L.~Barack {\em et~al.}, ``{Black holes, gravitational waves and fundamental
  physics: a roadmap},''
\href{http://arxiv.org/abs/1806.05195}{{\ttfamily arXiv:1806.05195 [gr-qc]}}.

\bibitem{Berti:2018vdi}
E.~Berti, K.~Yagi, H.~Yang, and N.~Yunes, ``{Extreme Gravity Tests with
  Gravitational Waves from Compact Binary Coalescences: (II) Ringdown},''
  \href{http://dx.doi.org/10.1007/s10714-018-2372-6}{{\em Gen. Rel. Grav.}
  {\bfseries 50} no.~5, (2018) 49},
\href{http://arxiv.org/abs/1801.03587}{{\ttfamily arXiv:1801.03587 [gr-qc]}}.

\bibitem{Berti:2019xgr}
E.~Berti {\em et~al.}, ``{Tests of General Relativity and Fundamental Physics
  with Space-based Gravitational Wave Detectors},''
\href{http://arxiv.org/abs/1903.02781}{{\ttfamily arXiv:1903.02781
  [astro-ph.HE]}}.

\bibitem{Sathyaprakash:2019yqt}
B.~S. Sathyaprakash {\em et~al.}, ``{Extreme Gravity and Fundamental
  Physics},''
\href{http://arxiv.org/abs/1903.09221}{{\ttfamily arXiv:1903.09221
  [astro-ph.HE]}}.

\bibitem{Carter71}
B.~Carter, ``Axisymmetric black hole has only two degrees of freedom,''
  \href{http://dx.doi.org/10.1103/PhysRevLett.26.331}{{\em Phys. Rev. Lett.}
  {\bfseries 26} (Feb, 1971) 331--333}.
  \url{http://link.aps.org/doi/10.1103/PhysRevLett.26.331}.

\bibitem{Hawking:1973uf}
S.~W. Hawking and G.~F.~R. Ellis, {\em {The Large Scale Structure of
  Space-Time}}.
\newblock Cambridge Monographs on Mathematical Physics. Cambridge University
  Press,
2011.
\newblock

\bibitem{Robinson}
D.~Robinson, {\em {Four decades of black holes uniqueness theorems}}.
\newblock Cambridge University Press, 2009.

\bibitem{Cardoso:2016ryw}
V.~Cardoso and L.~Gualtieri, ``{Testing the black hole ``no-hair''
  hypothesis},'' \href{http://dx.doi.org/10.1088/0264-9381/33/17/174001}{{\em
  Class. Quant. Grav.} {\bfseries 33} no.~17, (2016) 174001},
\href{http://arxiv.org/abs/1607.03133}{{\ttfamily arXiv:1607.03133 [gr-qc]}}.

\bibitem{Berti:2007zu}
E.~Berti, J.~Cardoso, V.~Cardoso, and M.~Cavaglia, ``{Matched-filtering and
  parameter estimation of ringdown waveforms},''
  \href{http://dx.doi.org/10.1103/PhysRevD.76.104044}{{\em Phys. Rev.}
  {\bfseries D76} (2007) 104044},
\href{http://arxiv.org/abs/0707.1202}{{\ttfamily arXiv:0707.1202 [gr-qc]}}.

\bibitem{Gossan:2011ha}
S.~Gossan, J.~Veitch, and B.~S. Sathyaprakash, ``{Bayesian model selection for
  testing the no-hair theorem with black hole ringdowns},''
  \href{http://dx.doi.org/10.1103/PhysRevD.85.124056}{{\em Phys. Rev.}
  {\bfseries D85} (2012) 124056},
\href{http://arxiv.org/abs/1111.5819}{{\ttfamily arXiv:1111.5819 [gr-qc]}}.

\bibitem{Meidam:2014jpa}
J.~Meidam, M.~Agathos, C.~Van Den~Broeck, J.~Veitch, and B.~S. Sathyaprakash,
  ``{Testing the no-hair theorem with black hole ringdowns using TIGER},''
  \href{http://dx.doi.org/10.1103/PhysRevD.90.064009}{{\em Phys. Rev.}
  {\bfseries D90} no.~6, (2014) 064009},
\href{http://arxiv.org/abs/1406.3201}{{\ttfamily arXiv:1406.3201 [gr-qc]}}.

\bibitem{Bhagwat:2017tkm}
S.~Bhagwat, M.~Okounkova, S.~W. Ballmer, D.~A. Brown, M.~Giesler, M.~A. Scheel,
  and S.~A. Teukolsky, ``{On choosing the start time of binary black hole
  ringdowns},'' \href{http://dx.doi.org/10.1103/PhysRevD.97.104065}{{\em Phys.
  Rev.} {\bfseries D97} no.~10, (2018) 104065},
\href{http://arxiv.org/abs/1711.00926}{{\ttfamily arXiv:1711.00926 [gr-qc]}}.

\bibitem{Baibhav:2017jhs}
V.~Baibhav, E.~Berti, V.~Cardoso, and G.~Khanna, ``{Black Hole Spectroscopy:
  Systematic Errors and Ringdown Energy Estimates},''
  \href{http://dx.doi.org/10.1103/PhysRevD.97.044048}{{\em Phys. Rev.}
  {\bfseries D97} no.~4, (2018) 044048},
\href{http://arxiv.org/abs/1710.02156}{{\ttfamily arXiv:1710.02156 [gr-qc]}}.

\bibitem{Baibhav:2018rfk}
V.~Baibhav and E.~Berti, ``{Multimode black hole spectroscopy},''
  \href{http://dx.doi.org/10.1103/PhysRevD.99.024005}{{\em Phys. Rev.}
  {\bfseries D99} no.~2, (2019) 024005},
\href{http://arxiv.org/abs/1809.03500}{{\ttfamily arXiv:1809.03500 [gr-qc]}}.

\bibitem{Brito:2018rfr}
R.~Brito, A.~Buonanno, and V.~Raymond, ``{Black-hole Spectroscopy by Making
  Full Use of Gravitational-Wave Modeling},''
  \href{http://dx.doi.org/10.1103/PhysRevD.98.084038}{{\em Phys. Rev.}
  {\bfseries D98} no.~8, (2018) 084038},
\href{http://arxiv.org/abs/1805.00293}{{\ttfamily arXiv:1805.00293 [gr-qc]}}.

\bibitem{Carullo:2018sfu}
G.~Carullo {\em et~al.}, ``{Empirical tests of the black hole no-hair
  conjecture using gravitational-wave observations},''
  \href{http://dx.doi.org/10.1103/PhysRevD.98.104020}{{\em Phys. Rev.}
  {\bfseries D98} no.~10, (2018) 104020},
\href{http://arxiv.org/abs/1805.04760}{{\ttfamily arXiv:1805.04760 [gr-qc]}}.

\bibitem{Hughes:2019zmt}
S.~A. Hughes, A.~Apte, G.~Khanna, and H.~Lim, ``{Learning about black hole
  binaries from their ringdown spectra},''
  \href{http://dx.doi.org/10.1103/PhysRevLett.123.161101}{{\em Phys. Rev.
  Lett.} {\bfseries 123} no.~16, (2019) 161101},
\href{http://arxiv.org/abs/1901.05900}{{\ttfamily arXiv:1901.05900 [gr-qc]}}.

\bibitem{Apte:2019txp}
A.~Apte and S.~A. Hughes, ``{Exciting black hole modes via misaligned
  coalescences: I. Inspiral, transition, and plunge trajectories using a
  generalized Ori-Thorne procedure},''
  \href{http://dx.doi.org/10.1103/PhysRevD.100.084031}{{\em Phys. Rev.}
  {\bfseries D100} no.~8, (2019) 084031},
\href{http://arxiv.org/abs/1901.05901}{{\ttfamily arXiv:1901.05901 [gr-qc]}}.

\bibitem{Lim:2019xrb}
H.~Lim, G.~Khanna, A.~Apte, and S.~A. Hughes, ``{Exciting black hole modes via
  misaligned coalescences: II. The mode content of late-time coalescence
  waveforms},'' \href{http://dx.doi.org/10.1103/PhysRevD.100.084032}{{\em Phys.
  Rev.} {\bfseries D100} no.~8, (2019) 084032},
\href{http://arxiv.org/abs/1901.05902}{{\ttfamily arXiv:1901.05902 [gr-qc]}}.

\bibitem{Audley:2017drz}
H.~{Audley}, S.~{Babak}, J.~{Baker}, E.~{Barausse}, P.~{Bender}, E.~{Berti},
  P.~{Binetruy}, M.~{Born}, D.~{Bortoluzzi}, J.~{Camp}, C.~{Caprini},
  V.~{Cardoso}, M.~{Colpi}, J.~{Conklin}, N.~{Cornish}, C.~{Cutler}, {\em
  et~al.}, ``{Laser Interferometer Space Antenna},'' {\em ArXiv e-prints}
  (Feb., 2017) , \href{http://arxiv.org/abs/1702.00786}{{\ttfamily
  arXiv:1702.00786 [astro-ph.IM]}}.

\bibitem{Punturo:2010zz}
M.~Punturo {\em et~al.}, ``{The Einstein Telescope: A third-generation
  gravitational wave observatory},''
\href{http://dx.doi.org/10.1088/0264-9381/27/19/194002}{{\em Class. Quant.
  Grav.} {\bfseries 27} (2010) 194002}.

\bibitem{Evans:2016mbw}
{\bfseries LIGO Scientific} Collaboration, B.~P. Abbott {\em et~al.},
  ``{Exploring the Sensitivity of Next Generation Gravitational Wave
  Detectors},'' \href{http://dx.doi.org/10.1088/1361-6382/aa51f4}{{\em Class.
  Quant. Grav.} {\bfseries 34} no.~4, (2017) 044001},
\href{http://arxiv.org/abs/1607.08697}{{\ttfamily arXiv:1607.08697
  [astro-ph.IM]}}.

\bibitem{Berti:2016lat}
E.~Berti, A.~Sesana, E.~Barausse, V.~Cardoso, and K.~Belczynski,
  ``{Spectroscopy of Kerr black holes with Earth- and space-based
  interferometers},''
  \href{http://dx.doi.org/10.1103/PhysRevLett.117.101102}{{\em Phys. Rev.
  Lett.} {\bfseries 117} no.~10, (2016) 101102},
\href{http://arxiv.org/abs/1605.09286}{{\ttfamily arXiv:1605.09286 [gr-qc]}}.

\bibitem{TheLIGOScientific:2016src}
{\bfseries Virgo, LIGO Scientific} Collaboration, B.~P. Abbott {\em et~al.},
  ``{Tests of general relativity with GW150914},''
  \href{http://dx.doi.org/10.1103/PhysRevLett.116.221101}{{\em Phys. Rev.
  Lett.} {\bfseries 116} no.~22, (2016) 221101},
\href{http://arxiv.org/abs/1602.03841}{{\ttfamily arXiv:1602.03841 [gr-qc]}}.

\bibitem{LIGOScientific:2019fpa}
{\bfseries LIGO Scientific, Virgo} Collaboration, B.~P. Abbott {\em et~al.},
  ``{Tests of General Relativity with the Binary Black Hole Signals from the
  LIGO-Virgo Catalog GWTC-1},''
\href{http://arxiv.org/abs/1903.04467}{{\ttfamily arXiv:1903.04467 [gr-qc]}}.

\bibitem{Giesler:2019uxc}
M.~Giesler, M.~Isi, M.~Scheel, and S.~Teukolsky, ``{Black hole ringdown: the
  importance of overtones},''
\href{http://arxiv.org/abs/1903.08284}{{\ttfamily arXiv:1903.08284 [gr-qc]}}.

\bibitem{Isi:2019aib}
M.~Isi, M.~Giesler, W.~M. Farr, M.~A. Scheel, and S.~A. Teukolsky, ``{Testing
  the no-hair theorem with GW150914},''
\href{http://arxiv.org/abs/1905.00869}{{\ttfamily arXiv:1905.00869 [gr-qc]}}.

\bibitem{Bhagwat:2019dtm}
S.~Bhagwat, X.~J. Forteza, P.~Pani, and V.~Ferrari, ``{Ringdown overtones,
  black hole spectroscopy and, no-hair theorem tests},''
\href{http://arxiv.org/abs/1910.08708}{{\ttfamily arXiv:1910.08708 [gr-qc]}}.

\bibitem{Barausse:2014tra}
E.~Barausse, V.~Cardoso, and P.~Pani, ``{Can environmental effects spoil
  precision gravitational-wave astrophysics?},''
  \href{http://dx.doi.org/10.1103/PhysRevD.89.104059}{{\em Phys. Rev.}
  {\bfseries D89} no.~10, (2014) 104059},
\href{http://arxiv.org/abs/1404.7149}{{\ttfamily arXiv:1404.7149 [gr-qc]}}.

\bibitem{Glampedakis:2017dvb}
K.~Glampedakis, G.~Pappas, H.~O. Silva, and E.~Berti, ``{Post-Kerr black hole
  spectroscopy},'' \href{http://dx.doi.org/10.1103/PhysRevD.96.064054}{{\em
  Phys. Rev.} {\bfseries D96} no.~6, (2017) 064054},
\href{http://arxiv.org/abs/1706.07658}{{\ttfamily arXiv:1706.07658 [gr-qc]}}.

\bibitem{Glampedakis:2017cgd}
K.~Glampedakis and G.~Pappas, ``{How well can ultracompact bodies imitate black
  hole ringdowns?},'' \href{http://dx.doi.org/10.1103/PhysRevD.97.041502}{{\em
  Phys. Rev.} {\bfseries D97} no.~4, (2018) 041502},
\href{http://arxiv.org/abs/1710.02136}{{\ttfamily arXiv:1710.02136 [gr-qc]}}.

\bibitem{Tattersall:2017erk}
O.~J. Tattersall, P.~G. Ferreira, and M.~Lagos, ``{General theories of linear
  gravitational perturbations to a Schwarzschild Black Hole},''
  \href{http://dx.doi.org/10.1103/PhysRevD.97.044021}{{\em Phys. Rev.}
  {\bfseries D97} no.~4, (2018) 044021},
\href{http://arxiv.org/abs/1711.01992}{{\ttfamily arXiv:1711.01992 [gr-qc]}}.

\bibitem{Franciolini:2018uyq}
G.~Franciolini, L.~Hui, R.~Penco, L.~Santoni, and E.~Trincherini, ``{Effective
  Field Theory of Black Hole Quasinormal Modes in Scalar-Tensor Theories},''
  \href{http://dx.doi.org/10.1007/JHEP02(2019)127}{{\em JHEP} {\bfseries 02}
  (2019) 127},
\href{http://arxiv.org/abs/1810.07706}{{\ttfamily arXiv:1810.07706 [hep-th]}}.

\bibitem{Cardoso:2019mqo}
V.~Cardoso, M.~Kimura, A.~Maselli, E.~Berti, C.~F.~B. Macedo, and R.~McManus,
  ``{Parametrized black hole quasinormal ringdown. I. Decoupled equations for
  nonrotating black holes},''
\href{http://arxiv.org/abs/1901.01265}{{\ttfamily arXiv:1901.01265 [gr-qc]}}.

\bibitem{McManus:2019ulj}
R.~McManus, E.~Berti, C.~F.~B. Macedo, M.~Kimura, A.~Maselli, and V.~Cardoso,
  ``{Parametrized black hole quasinormal ringdown. II. Coupled equations and
  quadratic corrections for nonrotating black holes},''
  \href{http://dx.doi.org/10.1103/PhysRevD.100.044061}{{\em Phys. Rev.}
  {\bfseries D100} no.~4, (2019) 044061},
\href{http://arxiv.org/abs/1906.05155}{{\ttfamily arXiv:1906.05155 [gr-qc]}}.

\bibitem{Glampedakis:2019dqh}
K.~Glampedakis and H.~O. Silva, ``{Eikonal quasinormal modes of black holes
  beyond General Relativity},''
  \href{http://dx.doi.org/10.1103/PhysRevD.100.044040}{{\em Phys. Rev.}
  {\bfseries D100} no.~4, (2019) 044040},
\href{http://arxiv.org/abs/1906.05455}{{\ttfamily arXiv:1906.05455 [gr-qc]}}.

\bibitem{Buonanno:2006ui}
A.~Buonanno, G.~B. Cook, and F.~Pretorius, ``{Inspiral, merger and ring-down of
  equal-mass black-hole binaries},''
  \href{http://dx.doi.org/10.1103/PhysRevD.75.124018}{{\em Phys. Rev.}
  {\bfseries D75} (2007) 124018},
\href{http://arxiv.org/abs/gr-qc/0610122}{{\ttfamily arXiv:gr-qc/0610122
  [gr-qc]}}.

\bibitem{Berti:2007fi}
E.~Berti, V.~Cardoso, J.~A. Gonzalez, U.~Sperhake, M.~Hannam, S.~Husa, and
  B.~Bruegmann, ``{Inspiral, merger and ringdown of unequal mass black hole
  binaries: A Multipolar analysis},''
  \href{http://dx.doi.org/10.1103/PhysRevD.76.064034}{{\em Phys. Rev.}
  {\bfseries D76} (2007) 064034},
\href{http://arxiv.org/abs/gr-qc/0703053}{{\ttfamily arXiv:gr-qc/0703053
  [GR-QC]}}.

\bibitem{Berti:2008af}
E.~Berti and M.~Volonteri, ``{Cosmological black hole spin evolution by mergers
  and accretion},'' \href{http://dx.doi.org/10.1086/590379}{{\em Astrophys. J.}
  {\bfseries 684} (2008) 822--828},
\href{http://arxiv.org/abs/0802.0025}{{\ttfamily arXiv:0802.0025 [astro-ph]}}.

\bibitem{Hofmann:2016yih}
F.~Hofmann, E.~Barausse, and L.~Rezzolla, ``{The final spin from binary black
  holes in quasi-circular orbits},''
  \href{http://dx.doi.org/10.3847/2041-8205/825/2/L19}{{\em Astrophys. J.}
  {\bfseries 825} no.~2, (2016) L19},
\href{http://arxiv.org/abs/1605.01938}{{\ttfamily arXiv:1605.01938 [gr-qc]}}.

\bibitem{Pani:2011gy}
P.~Pani, C.~F.~B. Macedo, L.~C.~B. Crispino, and V.~Cardoso, ``{Slowly rotating
  black holes in alternative theories of gravity},''
  \href{http://dx.doi.org/10.1103/PhysRevD.84.087501}{{\em Phys. Rev.}
  {\bfseries D84} (2011) 087501},
\href{http://arxiv.org/abs/1109.3996}{{\ttfamily arXiv:1109.3996 [gr-qc]}}.

\bibitem{Kleihaus:2011tg}
B.~Kleihaus, J.~Kunz, and E.~Radu, ``{Rotating Black Holes in Dilatonic
  Einstein-Gauss-Bonnet Theory},''
  \href{http://dx.doi.org/10.1103/PhysRevLett.106.151104}{{\em Phys. Rev.
  Lett.} {\bfseries 106} (2011) 151104},
\href{http://arxiv.org/abs/1101.2868}{{\ttfamily arXiv:1101.2868 [gr-qc]}}.

\bibitem{Ayzenberg:2014aka}
D.~Ayzenberg and N.~Yunes, ``{Slowly-Rotating Black Holes in
  Einstein-Dilaton-Gauss-Bonnet Gravity: Quadratic Order in Spin Solutions},''
  \href{http://dx.doi.org/10.1103/PhysRevD.91.069905,
  10.1103/PhysRevD.90.044066}{{\em Phys. Rev.} {\bfseries D90} (2014) 044066},
  \href{http://arxiv.org/abs/1405.2133}{{\ttfamily arXiv:1405.2133 [gr-qc]}}.
[Erratum: Phys. Rev.D91,no.6,069905(2015)].

\bibitem{Maselli:2015tta}
A.~Maselli, P.~Pani, L.~Gualtieri, and V.~Ferrari, ``{Rotating black holes in
  Einstein-Dilaton-Gauss-Bonnet gravity with finite coupling},''
  \href{http://dx.doi.org/10.1103/PhysRevD.92.083014}{{\em Phys. Rev.}
  {\bfseries D92} no.~8, (2015) 083014},
\href{http://arxiv.org/abs/1507.00680}{{\ttfamily arXiv:1507.00680 [gr-qc]}}.

\bibitem{Barausse:2015frm}
E.~Barausse, T.~P. Sotiriou, and I.~Vega, ``{Slowly rotating black holes in
  Einstein-æther theory},''
  \href{http://dx.doi.org/10.1103/PhysRevD.93.044044}{{\em Phys. Rev.}
  {\bfseries D93} no.~4, (2016) 044044},
\href{http://arxiv.org/abs/1512.05894}{{\ttfamily arXiv:1512.05894 [gr-qc]}}.

\bibitem{Herdeiro:2016tmi}
C.~Herdeiro, E.~Radu, and H.~Runarsson, ``{Kerr black holes with Proca hair},''
  \href{http://dx.doi.org/10.1088/0264-9381/33/15/154001}{{\em Class. Quant.
  Grav.} {\bfseries 33} no.~15, (2016) 154001},
\href{http://arxiv.org/abs/1603.02687}{{\ttfamily arXiv:1603.02687 [gr-qc]}}.

\bibitem{Cunha:2019dwb}
P.~V.~P. Cunha, C.~A.~R. Herdeiro, and E.~Radu, ``{Spontaneously scalarised
  Kerr black holes},''
\href{http://arxiv.org/abs/1904.09997}{{\ttfamily arXiv:1904.09997 [gr-qc]}}.

\bibitem{Herdeiro:2015waa}
C.~A.~R. Herdeiro and E.~Radu, ``{Asymptotically flat black holes with scalar
  hair: a review},'' \href{http://dx.doi.org/10.1142/S0218271815420146}{{\em
  Int. J. Mod. Phys.} {\bfseries D24} no.~09, (2015) 1542014},
\href{http://arxiv.org/abs/1504.08209}{{\ttfamily arXiv:1504.08209 [gr-qc]}}.

\bibitem{Yagi:2016jml}
K.~Yagi and L.~C. Stein, ``{Black Hole Based Tests of General Relativity},''
  \href{http://dx.doi.org/10.1088/0264-9381/33/5/054001}{{\em Class. Quant.
  Grav.} {\bfseries 33} no.~5, (2016) 054001},
\href{http://arxiv.org/abs/1602.02413}{{\ttfamily arXiv:1602.02413 [gr-qc]}}.

\bibitem{Teukolsky:1972my}
S.~A. Teukolsky, ``{Rotating black holes - separable wave equations for
  gravitational and electromagnetic perturbations},''
\href{http://dx.doi.org/10.1103/PhysRevLett.29.1114}{{\em Phys. Rev. Lett.}
  {\bfseries 29} (1972) 1114--1118}.

\bibitem{Teukolsky:1973ha}
S.~A. Teukolsky, ``{Perturbations of a rotating black hole. 1. Fundamental
  equations for gravitational electromagnetic and neutrino field
  perturbations},''
\href{http://dx.doi.org/10.1086/152444}{{\em Astrophys.J.} {\bfseries 185}
  (1973) 635--647}.

\bibitem{Pani:2013pma}
P.~Pani, ``{Advanced Methods in Black-Hole Perturbation Theory},''
  \href{http://dx.doi.org/10.1142/S0217751X13400186}{{\em Int. J. Mod. Phys.}
  {\bfseries A28} (2013) 1340018},
\href{http://arxiv.org/abs/1305.6759}{{\ttfamily arXiv:1305.6759 [gr-qc]}}.

\bibitem{Dias:2015wqa}
O.~J.~C. Dias, M.~Godazgar, and J.~E. Santos, ``{Linear Mode Stability of the
  Kerr-Newman Black Hole and Its Quasinormal Modes},''
  \href{http://dx.doi.org/10.1103/PhysRevLett.114.151101}{{\em Phys. Rev.
  Lett.} {\bfseries 114} no.~15, (2015) 151101},
\href{http://arxiv.org/abs/1501.04625}{{\ttfamily arXiv:1501.04625 [gr-qc]}}.

\bibitem{Okounkova:2017yby}
M.~Okounkova, L.~C. Stein, M.~A. Scheel, and D.~A. Hemberger, ``{Numerical
  binary black hole mergers in dynamical Chern-Simons gravity: Scalar field},''
  \href{http://dx.doi.org/10.1103/PhysRevD.96.044020}{{\em Phys. Rev.}
  {\bfseries D96} no.~4, (2017) 044020},
\href{http://arxiv.org/abs/1705.07924}{{\ttfamily arXiv:1705.07924 [gr-qc]}}.

\bibitem{Witek:2018dmd}
H.~Witek, L.~Gualtieri, P.~Pani, and T.~P. Sotiriou, ``{Black holes and binary
  mergers in scalar Gauss-Bonnet gravity: scalar field dynamics},''
\href{http://arxiv.org/abs/1810.05177}{{\ttfamily arXiv:1810.05177 [gr-qc]}}.

\bibitem{Okounkova:2019dfo}
M.~Okounkova, L.~C. Stein, M.~A. Scheel, and S.~A. Teukolsky, ``{Numerical
  binary black hole collisions in dynamical Chern-Simons gravity},''
\href{http://arxiv.org/abs/1906.08789}{{\ttfamily arXiv:1906.08789 [gr-qc]}}.

\bibitem{Okounkova:2019zep}
M.~Okounkova, ``{Stability of rotating black holes in Einstein dilaton
  Gauss-Bonnet gravity},''
\href{http://arxiv.org/abs/1909.12251}{{\ttfamily arXiv:1909.12251 [gr-qc]}}.

\bibitem{Cardoso:2018ptl}
V.~Cardoso, M.~Kimura, A.~Maselli, and L.~Senatore, ``{Black Holes in an
  Effective Field Theory Extension of General Relativity},''
  \href{http://dx.doi.org/10.1103/PhysRevLett.121.251105}{{\em Phys. Rev.
  Lett.} {\bfseries 121} no.~25, (2018) 251105},
\href{http://arxiv.org/abs/1808.08962}{{\ttfamily arXiv:1808.08962 [gr-qc]}}.

\bibitem{Barausse:2008xv}
E.~Barausse and T.~P. Sotiriou, ``{Perturbed Kerr Black Holes can probe
  deviations from General Relativity},''
  \href{http://dx.doi.org/10.1103/PhysRevLett.101.099001}{{\em Phys. Rev.
  Lett.} {\bfseries 101} (2008) 099001},
\href{http://arxiv.org/abs/0803.3433}{{\ttfamily arXiv:0803.3433 [gr-qc]}}.

\bibitem{Molina:2010fb}
C.~Molina, P.~Pani, V.~Cardoso, and L.~Gualtieri, ``{Gravitational signature of
  Schwarzschild black holes in dynamical Chern-Simons gravity},''
  \href{http://dx.doi.org/10.1103/PhysRevD.81.124021}{{\em Phys. Rev.}
  {\bfseries D81} (2010) 124021},
\href{http://arxiv.org/abs/1004.4007}{{\ttfamily arXiv:1004.4007 [gr-qc]}}.

\bibitem{Tattersall:2018map}
O.~J. Tattersall, P.~G. Ferreira, and M.~Lagos, ``{Speed of gravitational waves
  and black hole hair},''
  \href{http://dx.doi.org/10.1103/PhysRevD.97.084005}{{\em Phys. Rev.}
  {\bfseries D97} no.~8, (2018) 084005},
\href{http://arxiv.org/abs/1802.08606}{{\ttfamily arXiv:1802.08606 [gr-qc]}}.

\bibitem{Tattersall:2018nve}
O.~J. Tattersall and P.~G. Ferreira, ``{Quasinormal modes of black holes in
  Horndeski gravity},''
  \href{http://dx.doi.org/10.1103/PhysRevD.97.104047}{{\em Phys. Rev.}
  {\bfseries D97} no.~10, (2018) 104047},
\href{http://arxiv.org/abs/1804.08950}{{\ttfamily arXiv:1804.08950 [gr-qc]}}.

\bibitem{Tattersall:2019pvx}
O.~J. Tattersall and P.~G. Ferreira, ``{Forecasts for Low Spin Black Hole
  Spectroscopy in Horndeski Gravity},''
\href{http://arxiv.org/abs/1904.05112}{{\ttfamily arXiv:1904.05112 [gr-qc]}}.

\bibitem{Blazquez-Salcedo:2016enn}
J.~L. Blázquez-Salcedo, C.~F.~B. Macedo, V.~Cardoso, V.~Ferrari, L.~Gualtieri,
  F.~S. Khoo, J.~Kunz, and P.~Pani, ``{Perturbed black holes in
  Einstein-dilaton-Gauss-Bonnet gravity: Stability, ringdown, and
  gravitational-wave emission},''
  \href{http://dx.doi.org/10.1103/PhysRevD.94.104024}{{\em Phys. Rev.}
  {\bfseries D94} no.~10, (2016) 104024},
\href{http://arxiv.org/abs/1609.01286}{{\ttfamily arXiv:1609.01286 [gr-qc]}}.

\bibitem{Will:2014kxa}
C.~M. Will, ``{The Confrontation between General Relativity and Experiment},''
  \href{http://dx.doi.org/10.12942/lrr-2014-4}{{\em Living Rev. Rel.}
  {\bfseries 17} (2014) 4},
\href{http://arxiv.org/abs/1403.7377}{{\ttfamily arXiv:1403.7377 [gr-qc]}}.

\bibitem{Yunes:2009ke}
N.~Yunes and F.~Pretorius, ``{Fundamental Theoretical Bias in Gravitational
  Wave Astrophysics and the Parameterized Post-Einsteinian Framework},''
  \href{http://dx.doi.org/10.1103/PhysRevD.80.122003}{{\em Phys. Rev.}
  {\bfseries D80} (2009) 122003},
\href{http://arxiv.org/abs/0909.3328}{{\ttfamily arXiv:0909.3328 [gr-qc]}}.

\bibitem{Agathos:2013upa}
M.~Agathos, W.~Del~Pozzo, T.~G.~F. Li, C.~Van Den~Broeck, J.~Veitch, and
  S.~Vitale, ``{TIGER: A data analysis pipeline for testing the strong-field
  dynamics of general relativity with gravitational wave signals from
  coalescing compact binaries},''
  \href{http://dx.doi.org/10.1103/PhysRevD.89.082001}{{\em Phys. Rev.}
  {\bfseries D89} no.~8, (2014) 082001},
\href{http://arxiv.org/abs/1311.0420}{{\ttfamily arXiv:1311.0420 [gr-qc]}}.

\bibitem{Isi:2019asy}
M.~Isi, K.~Chatziioannou, and W.~M. Farr, ``{A hierarchical test of general
  relativity with gravitational waves},''
\href{http://arxiv.org/abs/1904.08011}{{\ttfamily arXiv:1904.08011 [gr-qc]}}.

\bibitem{Yang:2017zxs}
H.~Yang, K.~Yagi, J.~Blackman, L.~Lehner, V.~Paschalidis, F.~Pretorius, and
  N.~Yunes, ``{Black hole spectroscopy with coherent mode stacking},''
  \href{http://dx.doi.org/10.1103/PhysRevLett.118.161101}{{\em Phys. Rev.
  Lett.} {\bfseries 118} no.~16, (2017) 161101},
\href{http://arxiv.org/abs/1701.05808}{{\ttfamily arXiv:1701.05808 [gr-qc]}}.

\bibitem{Leaver:1986gd}
E.~W. Leaver, ``{Spectral decomposition of the perturbation response of the
  Schwarzschild geometry},''
\href{http://dx.doi.org/10.1103/PhysRevD.34.384}{{\em Phys. Rev.} {\bfseries
  D34} (1986) 384--408}.

\bibitem{Berti:2006wq}
E.~Berti and V.~Cardoso, ``{Quasinormal ringing of Kerr black holes. I. The
  Excitation factors},''
  \href{http://dx.doi.org/10.1103/PhysRevD.74.104020}{{\em Phys. Rev.}
  {\bfseries D74} (2006) 104020},
\href{http://arxiv.org/abs/gr-qc/0605118}{{\ttfamily arXiv:gr-qc/0605118
  [gr-qc]}}.

\bibitem{London:2014cma}
L.~London, D.~Shoemaker, and J.~Healy, ``{Modeling ringdown: Beyond the
  fundamental quasinormal modes},''
  \href{http://dx.doi.org/10.1103/PhysRevD.90.124032,
  10.1103/PhysRevD.94.069902}{{\em Phys. Rev.} {\bfseries D90} no.~12, (2014)
  124032}, \href{http://arxiv.org/abs/1404.3197}{{\ttfamily arXiv:1404.3197
  [gr-qc]}}.
[Erratum: Phys. Rev.D94,no.6,069902(2016)].

\bibitem{ringdown}
 \noindent\url{https://pages.jh.edu/~eberti2/ringdown/}\\\url{https://centra.tecnico.ulisboa.pt/~vitor/?page=ringdown}.

\bibitem{EspositoFarese:2003ze}
G.~Esposito-Farese, ``{Scalar tensor theories and cosmology and tests of a
  quintessence Gauss-Bonnet coupling},'' in {\em {38th Rencontres de Moriond on
  Gravitational Waves and Experimental Gravity Les Arcs, Savoie, France, March
  22-29, 2003}}.
\newblock 2003.
\newblock
\href{http://arxiv.org/abs/gr-qc/0306018}{{\ttfamily arXiv:gr-qc/0306018
  [gr-qc]}}.
\newblock

\bibitem{Barausse:2013nwa}
E.~Barausse and T.~P. Sotiriou, ``{Black holes in Lorentz-violating gravity
  theories},'' \href{http://dx.doi.org/10.1088/0264-9381/30/24/244010}{{\em
  Class. Quant. Grav.} {\bfseries 30} (2013) 244010},
\href{http://arxiv.org/abs/1307.3359}{{\ttfamily arXiv:1307.3359 [gr-qc]}}.

\bibitem{Mignemi:1992nt}
S.~Mignemi and N.~R. Stewart, ``{Charged black holes in effective string
  theory},'' \href{http://dx.doi.org/10.1103/PhysRevD.47.5259}{{\em Phys. Rev.}
  {\bfseries D47} (1993) 5259--5269},
\href{http://arxiv.org/abs/hep-th/9212146}{{\ttfamily arXiv:hep-th/9212146
  [hep-th]}}.

\bibitem{Yunes:2011we}
N.~Yunes and L.~C. Stein, ``{Non-Spinning Black Holes in Alternative Theories
  of Gravity},'' \href{http://dx.doi.org/10.1103/PhysRevD.83.104002}{{\em Phys.
  Rev.} {\bfseries D83} (2011) 104002},
\href{http://arxiv.org/abs/1101.2921}{{\ttfamily arXiv:1101.2921 [gr-qc]}}.

\bibitem{Julie:2019sab}
F.-L. Julié and E.~Berti, ``{Post-Newtonian dynamics and black hole
  thermodynamics in Einstein-scalar-Gauss-Bonnet gravity},''
\href{http://arxiv.org/abs/1909.05258}{{\ttfamily arXiv:1909.05258 [gr-qc]}}.

\bibitem{Yunes:2009hc}
N.~Yunes and F.~Pretorius, ``{Dynamical Chern-Simons Modified Gravity. I.
  Spinning Black Holes in the Slow-Rotation Approximation},''
  \href{http://dx.doi.org/10.1103/PhysRevD.79.084043}{{\em Phys. Rev.}
  {\bfseries D79} (2009) 084043},
\href{http://arxiv.org/abs/0902.4669}{{\ttfamily arXiv:0902.4669 [gr-qc]}}.

\bibitem{Maselli:2017kic}
A.~Maselli, P.~Pani, R.~Cotesta, L.~Gualtieri, V.~Ferrari, and L.~Stella,
  ``{Geodesic models of quasi-periodic-oscillations as probes of quadratic
  gravity},'' \href{http://dx.doi.org/10.3847/1538-4357/aa72e2}{{\em Astrophys.
  J.} {\bfseries 843} no.~1, (2017) 25},
\href{http://arxiv.org/abs/1703.01472}{{\ttfamily arXiv:1703.01472
  [astro-ph.HE]}}.

\bibitem{Cardoso:2016olt}
V.~Cardoso, C.~F.~B. Macedo, P.~Pani, and V.~Ferrari, ``{Black holes and
  gravitational waves in models of minicharged dark matter},''
  \href{http://dx.doi.org/10.1088/1475-7516/2016/05/054}{{\em JCAP} {\bfseries
  1605} no.~05, (2016) 054},
\href{http://arxiv.org/abs/1604.07845}{{\ttfamily arXiv:1604.07845 [hep-ph]}}.

\bibitem{Healy:2014yta}
J.~Healy, C.~O. Lousto, and Y.~Zlochower, ``{Remnant mass, spin, and recoil
  from spin aligned black-hole binaries},''
  \href{http://dx.doi.org/10.1103/PhysRevD.90.104004}{{\em Phys. Rev.}
  {\bfseries D90} no.~10, (2014) 104004},
\href{http://arxiv.org/abs/1406.7295}{{\ttfamily arXiv:1406.7295 [gr-qc]}}.

\bibitem{Hild:2009ns}
S.~Hild, S.~Chelkowski, A.~Freise, J.~Franc, N.~Morgado, R.~Flaminio, and
  R.~DeSalvo, ``{A Xylophone Configuration for a third Generation Gravitational
  Wave Detector},'' \href{http://dx.doi.org/10.1088/0264-9381/27/1/015003}{{\em
  Class. Quant. Grav.} {\bfseries 27} (2010) 015003},
\href{http://arxiv.org/abs/0906.2655}{{\ttfamily arXiv:0906.2655 [gr-qc]}}.

\bibitem{Maselli:2017kvl}
A.~Maselli, K.~Kokkotas, and P.~Laguna, ``{Observing binary black hole
  ringdowns by advanced gravitational wave detectors},''
  \href{http://dx.doi.org/10.1103/PhysRevD.95.104026}{{\em Phys. Rev.}
  {\bfseries D95} no.~10, (2017) 104026},
\href{http://arxiv.org/abs/1702.01110}{{\ttfamily arXiv:1702.01110 [gr-qc]}}.

\bibitem{alLIGOScientificCollaborationandVirgoCollaboration:2018cn}
al~LIGO Scientific~Collaboration and B.~P. A.~e. Virgo~Collaboration,
  ``{GW170817: Implications for the Stochastic Gravitational-Wave Background
  from Compact Binary Coalescences},'' {\em Physical Review Letters} {\bfseries
  120} no.~9, (Feb., 2018) 091101.

\bibitem{Belczynski:2016jno}
K.~Belczynski {\em et~al.}, ``{The Effect of Pair-Instability Mass Loss on
  Black Hole Mergers},''
  \href{http://dx.doi.org/10.1051/0004-6361/201628980}{{\em Astron. Astrophys.}
  {\bfseries 594} (2016) A97},
\href{http://arxiv.org/abs/1607.03116}{{\ttfamily arXiv:1607.03116
  [astro-ph.HE]}}.

\bibitem{Berti:2006jo}
E.~Berti, V.~Cardoso, and C.~M. Will, ``{Gravitational-wave spectroscopy of
  massive black holes with the space interferometer LISA},'' {\em Physical
  Review D} {\bfseries 73} no.~6, (Mar., 2006) 2--41.

\bibitem{Gilks:1996}
W.~R. Gilks, S.~Richardson, and D.~J. Spiegelhalter, {\em {Markov Chain Monte
  Carlo in Practice}}.
\newblock Chapman \& Hall, London, UK, 1996.

\bibitem{1085030}
G.~Kjellstrom and L.~Taxen, ``Stochastic optimization in system design,''
  \href{http://dx.doi.org/10.1109/TCS.1981.1085030}{{\em IEEE Transactions on
  Circuits and Systems} {\bfseries 28} no.~7, (July, 1981) 702--715}.

\bibitem{5586491}
C.~L. Müller and I.~F. Sbalzarini, ``Gaussian adaptation as a unifying
  framework for continuous black-box optimization and adaptive monte carlo
  sampling,'' \href{http://dx.doi.org/10.1109/CEC.2010.5586491}{{\em IEEE
  Congress on Evolutionary Computation} (July, 2010) 1--8}.

\bibitem{Vallisneri:2013rc}
M.~Vallisneri and N.~Yunes, ``{Stealth Bias in Gravitational-Wave Parameter
  Estimation},'' \href{http://dx.doi.org/10.1103/PhysRevD.87.102002}{{\em Phys.
  Rev.} {\bfseries D87} no.~10, (2013) 102002},
\href{http://arxiv.org/abs/1301.2627}{{\ttfamily arXiv:1301.2627 [gr-qc]}}.

\end{thebibliography}\endgroup

\end{document}